\documentclass[amsmath,amssymb,aps,floats,amsfonts,notitlepage,superscriptaddress,eqsecnum,nofootinbib,prd,a4paper,longbibliography]{revtex4-1}
\allowdisplaybreaks

\usepackage[utf8]{inputenc}
\usepackage[]{graphicx}
\usepackage{hyperref}
\usepackage{url}
\usepackage{color}
\usepackage[usenames,dvipsnames,svgnames,table]{xcolor}
\usepackage{multirow}

\newcommand{\fixme}[1]{}
\newcommand{\MC}[1]{}
\newcommand{\CK}[1]{}
\newcommand{\AO}[1]{}

\newcommand{\indmode}{\ell m}

\newcommand{\RAM}{\nu}
\newcommand{\eps}{\epsilon}

\newcommand{\Disc}{\delta}
\newcommand{\Gret}{G_{\textit{ret}}}
\newcommand{\Slm}{ S_{\ell m\omega}}

\newcommand{\Rinhat}{\hat{R}^{\text{in}}_{\ell m}}
\newcommand{\Ruphat}{\hat{R}^{\text{up}}_{\ell m}}

\newcommand{\Rin}{R^{\text{in}}_{\ell m}}
\newcommand{\Rup}{R^{\text{up}}_{\ell m}}
\newcommand{\Rupp}{R^{\text{up},+}_{\ell m}}
\newcommand{\Rupm}{R^{\text{up}}_{\ell m,-}}
\newcommand{\Rupphat}{\hat{R}^{\text{up},+}_{\ell m}}
\newcommand{\Rupmhat}{\hat{R}^{\text{up},-}_{\ell m}}
\newcommand{\Ruppmhat}{\hat{R}^{\text{up},\pm}_{\ell m}}
\newcommand{\Ruppm}{R^{\text{up},\pm}_{\ell m}}
\newcommand{\Rinup}{R^{in/up}_{\indmode}}
\newcommand{\Ctrans}{C^{\text{tra}}}
\newcommand{\Btrans}{B^{\text{tra}}}
\newcommand{\Binc}{B^{\text{inc}}}
\newcommand{\Bref}{B^{\text{ref}}}
\newcommand{\Btransp}{B^{\text{tra},+}}
\newcommand{\Bincp}{B^{\text{inc},+}}
\newcommand{\Brefp}{B^{\text{ref},+}}
\newcommand{\nn}{\nonumber}
\newcommand{\an}[1]{a_{#1}}

\newcommand{\Wbp}{W^+}
\newcommand{\Wbm}{W^-}

\newcommand{\Glm}{G_{\indmode}}

\newcommand{\tort}{r_*}
\newcommand{\ph}{\text{ph}}

\newcommand{\fNIA}{\sigma}
\newcommand{\tfreq}{\tilde{\omega}}

\newcommand{\pdiff}[2]  {\frac{\partial #1}{\partial #2}}
\newcommand{\spdiff}[2] {\frac{\partial^2 #1}{\partial #2^2}}

\begin{document}

\title{High-order tail in Kerr spacetime}

\author{Marc Casals}
\email{mcasals@cbpf.br,marc.casals@ucd.ie.}
\affiliation{Centro Brasileiro de Pesquisas F\'isicas (CBPF), Rio de Janeiro, 
CEP 22290-180, 
Brazil.}
\affiliation{School of Mathematics and Statistics and Complex \& Adaptive Systems Laboratory, University College Dublin, Belfield, Dublin 4, Ireland.}

\author{Chris Kavanagh}
\email{christopher.kavanagh@ucdconnect.ie.}
\affiliation{School of Mathematics and Statistics and Complex \& Adaptive Systems Laboratory, University College Dublin, Belfield, Dublin 4, Ireland.}

\author{Adrian C.~Ottewill}
\email{adrian.ottewill@ucd.ie.}
\affiliation{School of Mathematics and Statistics and Complex \& Adaptive Systems Laboratory, University College Dublin, Belfield, Dublin 4, Ireland.}

\begin{abstract}

We investigate the late-time tail of the retarded Green function for the dynamics of a linear field perturbation of Kerr spacetime.
We develop an analytical formalism  for obtaining the late-time tail up to arbitrary order for general integer spin of the field.
We then apply this formalism to obtain
 the details of 
the first five orders in the late-time tail  of the Green function for the case of a scalar field:
to leading order we recover the known power law tail  $t^{-2\ell-3}$,
 and
 at third
  order we obtain a logarithmic correction, 
  $t^{-2\ell-5}\ln t$,
  \MC{To be checked}
where $\ell$ is the field multipole.

\end{abstract}

\maketitle


\section{Introduction}

The study of the mode-decomposition of linear field perturbations of spherical (Schwarzschild) and axially-symmetric (Kerr)
black hole spacetimes has a long history in the General Relativity literature.
Such study has been applied
to a broad range of astrophysical questions such as stability of black holes (e.g.,~\cite{Regge:1957td,PhysRevD.2.2141}
in Schwarzschild spacetime and~\cite{whiting1989mode,Hartle:Wilkins:1974,casals2016horizon} in Kerr spacetime), 
the self-force on a point particle moving on a curved background~\cite{Poisson:2011nh}
and
the end stages of gravitational collapse and of mergers of black holes.  
These end stages typically present an
exponentially-decaying `ringdown' in the field  (which was observed in the historical detection of gravitational waves by the Laser Interferometer gravitational-wave Observatory~\cite{PhysRevLett.116.061102}) followed by a late-time behaviour.

The late-time behaviour of scalar (spin-$0$), electromagnetic (spin-$1$) and gravitational (spin-$2$) field perturbations was first presented by Price \cite{Price:1971fb,Price:1972pw},
in the case of Schwarzschild spacetime. Price found that the multipole-$\ell$ field moments behave at late times (and fixed radius) as
a power law decay of $t_S^{-2\ell-3}$, where $t_S$ is the standard Schwarzschild time and $\ell$ is the field multipole. 
This has henceforth been referred to as the late-time power law-tail of the Schwarzschild black hole and can be interpreted astrophysically as the method with which a star undergoing spherically-symmetric gravitational collapse
settles down finally into a black hole with  `no hair' (i.e., its only  conserved charges are its mass and -- if it possesses any -- its
angular momentum and electrical
charge). 
The mathematics of this calculation was refined by Leaver \cite{Leaver:1986}
via an analysis in the complex-frequency domain of the retarded Green function of the wave equation satisfied by the field.
By deforming the Fourier-integration contour in the  complex-frequency plane, Leaver
 identified the source of the power law as coming from the branch cut that the Fourier modes of the Green function possess.
 In particular, he noted that it was the low-frequency asymptotics of the branch cut that gave the dominant
contribution to the Green function at late times. Using this insight into the nature of the decay tail, Hod calculated the leading-order branch cut contribution in Kerr spacetime finding the decay tail for  fields of spin-$0$, -$1$  and -$2$, all
at asymptotic null infinity, timelike infinity and at the event horizon \cite{PhysRevLett.84.10,hod2000mode}. 
The result at timelike infinity is that all scalar, electromagnetic and gravitational fields decay at late times
as $t^{-2\ell-3}$, 
where $t$ is  the Boyer-Lindquist coordinate and $\ell$ is the multipole number corresponding to a decomposition in spin-weighted spheroidal harmonics.
In a parallel series of works, Barack and Ori also calculated the leading order decay tail in Kerr spacetime:
in~\cite{barack1999late} for the scalar field at  null infinity, timelike infinity and on the event horizon, and in~\cite{PhysRevD.61.024026}
for the electromagnetic and gravitational fields on the event horizon. 
Their analysis however, was performed in the time domain, and in contrast to Hod's results, was valid at arbitrary radii. 
Recently, it has been observed~\cite{casals2016horizon} that an additional branch cut in the case of extreme Kerr gives rise to an instability  at late times of  the event horizon of this black hole, thus
generalizing  previous results by Aretakis for axisymmetric perturbations~\cite{Aretakis:2012ei,aretakis2012decay}.
Further to these works, there have also been many numerical investigations of the decay tails in Kerr spacetime in various asymptotic regimes. Much difficulty is encountered in these simulations due to different choices of co-ordinates and harmonic bases, see \cite{Zenginoglu:2012us} and references therein.

While the above results provide much physical insight into the nature of disturbances to the spacetime, they are not entirely sufficient for the applications of 
black hole perturbation theory to the 
calculation of the self-force.
The calculation of the self-force is important in order to model accurately the emission of gravitational waves by a black hole inspiral in the
extreme (or even intermediate~\cite{le2012gravitational}) mass-ratio regime (EMRIs and IMRIs).
In~\cite{CDOW13}, the self-force was calculated in the case of a scalar charge on Schwarschild spacetime by integrating 
the Green function over the past worldline of the charge. The Green function, in its turn, was calculated via Leaver's technique of contour-deformation
and it was observed that, for the branch cut contribution, the leading low-frequency asymptotics was not sufficient in order to obtain the self-force
accurately `enough'.
It was  discovered  that while the exponentially decaying
quasi-normal modes are dominant at intermediate times, the omission of the branch cut at these times (which comes from the branch cut modes to higher order in the frequency, as calculated in, e.g.,~\cite{PhysRevLett.109.111101,Casals:2012ng,Casals:2011aa,Casals:Ottewill:2015,casals2016quasi}) 
can lead to noticable errors, dispelling the identification of the branch cut with 
solely late times. A more accurate statement would be that the branch cut becomes \textit{dominant} at late times.

Combining Leaver's technique
 with the advances in analytic black hole perturbation theory 
provided by the 
method of Mano, Suzuki and Tagoshi (MST)
\cite{Mano:Suzuki:Takasugi:1996,Sasaki:2003xr}, two of us were able to calculate the higher
order corrections to Price's decay tail in Schwarzschild spacetime at arbitrary radii \cite{Casals:Ottewill:2015}. In calculating the corrections, it was 
shown
that the purely power-law nature, $t_S^{-2\ell-3}$, is `corrupted' by logarithmic terms at next-to-next-to-leading order, $t_S^{-2\ell-5}\ln\, t$. These terms allowed for a better approximation of the global Green function
in Schwarzschild spacetime needed for the past-history integral found in self force calculations, such as that in~\cite{CDOW13}.

In this paper we present an extension of the calculation of~\cite{Casals:Ottewill:2015} to (sub-extremal) Kerr spacetime.
We develop the MST method for the calculation of the branch cut contribution to the retarded Green function
 for field perturbations of general integer spin in Kerr.
 We then apply this formalism to calculate explicitly the late-time behaviour of a massless scalar field 
up  to five orders. 
We use Boyer-Lindquist time $t$.
Our leading order agrees with the literature results in Kerr, i.e., $t^{-2\ell-3}$.
We then find that a new logarithmic correction appears at next-to-next-to-leading order, i.e.,  $t^{-2\ell-5}\ln t$, as in Schwarzschild.
Finally, we compare our results with the fundamentally independent evaluation of the Green function via a real-frequency evaluation of the Fourier integral. 

The layout of the rest of this paper is as follows.
In Sec.\ref{sec:GF} we introduce the retarded Green function of the Teukolsky equation for spin-field perturbations of Kerr space-time.
In Sec.\ref{eq:MST} we summarize the main MST equations, already given in the literature, and which we need for later on.
In Sec.\ref{sec:GF compl} we introduce the deformation of the frequency-integral into the complex frequency plane.
In Sec.\ref{sec:BC} we develop the formalism for the branch cut integral and obtain analytical expressions for the Green function modes along the branch cut.
We obtain small-frequency expansions of the radius-independent part of these modes in Sec.\ref{sec:low freq r-indep}
and of the radial functions in Sec.\ref{sec:low freq r-dep}.
We put together these results in Sec.\ref{sec:late-time}, where we give the late-time tail of the Green function up to the first five orders.
We conclude the main body in Sec.\ref{sec:disc} with a discussion.
We have two appendices.
In App.\ref{sec:AngularCuts} we show that extra branch cuts that the angular eigenvalues and eigenfunctions have 
do not contribute to the Green function after summing over $\ell$.
In App.\ref{sec:MSTexpansions} we give small-frequency expansions for the series coefficients and for MST's so-called renormalized angular momentum parameter.
We choose units $c=G=1$ and, wherever ommited, $M=1$.


\section{Green function for the Teukolsky equation}\label{sec:GF}

The study of linear field perturbations $\psi$ on Kerr spacetime in Boyer-Lindquist coordinates $\{t,r,\theta,\phi\}$ and in the Kinnersley tetrad
 can be described in a unified way by the Teukolsky equation~\cite{Teukolsky:1973ha},
\begin{align}
	\left[\frac{(r^2+a^2)^2}{\Delta}-a^2\sin^2\theta\right]\spdiff{\psi}{t}+\frac{4 M a r}{\Delta}&\frac{\partial^2 \psi}{\partial t \partial \varphi}+
	\left[\frac{a^2}{\Delta}-\frac{1}{\sin^2\theta}\right]\spdiff{\psi}{\varphi} 
	-\Delta^{-s}\pdiff{}{r}\left(\Delta^{s+1}\pdiff{\psi}{r}\right) -\frac{1}{\sin \theta}\pdiff{}{\theta}\left(\sin\theta \pdiff{\psi}{\theta}\right)  \nonumber\\
	-2s\left[\frac{a(r-M)}{\Delta}+\frac{i \cos\theta}{\sin^2\theta}\right]&\pdiff{\psi}{\varphi}
	-2 s \left[\frac{M(r^2-a^2)}{\Delta}-r -i a \cos\theta\right]\pdiff{\psi}{t}+(s^2\cot^2\theta-s)\psi = 4 \pi \Sigma\cdot T,
	\label{Eq:TeukMaster}
\end{align}
where $\Delta\equiv r^2-2Mr+a^2=(r-r_+)(r-r_-)$, $\Sigma\equiv r^2+a^2 \cos^2\theta$, $T$ is the matter source term and $s$ denotes the spin of interest, $s=0$, $1$ and $2$ for scalar, electromagnetic and gravitational perturbations respectively
(the Teukolsky equation (\ref{Eq:TeukMaster}) is also valid for $s=1/2$ but we shall not consider this spin in this paper). 
The parameters $M$ and $a$ denote, respectively, the mass and angular momentum per unit mass.
Here, $r_{\pm}\equiv M\pm \sqrt{M^2-a^2}$ are the outer (event) horizon ($r_+$) and inner (Cauchy) horizon ($r_-$).
The Teukolsky equation (\ref{Eq:TeukMaster}) can be solved by calculating a Green function satisfying
\begin{align}
\mathcal{T}G(x,x')=4\pi\Sigma\cdot \delta_{4}(x,x'),
\end{align}	
where  $x$ and $x'$ are points in Kerr spacetime, $\mathcal{T}$ is the differential operator on the left hand side of Eq.(\ref{Eq:TeukMaster}), and $\delta_{4}(x,x')\equiv \delta_{4}(x-x')/\sqrt{|g|}$ is 
an invariant 4-dimensional dirac delta distribution, where $g=-\Sigma^2\sin^2\theta$ is the determinant of the metric.
Teukolsky also showed that, in the frequency domain, his equation can be separated into radial and angular components by
using the spin-weighted spheroidal harmonics. For the Green function this is achieved by writing
\begin{align}
G(x,x')=2\sum_{\ell=|s|}^{\infty}\sum_{m=-\ell}^{\ell}\int_{-\infty+i c}^{\infty+i c}d\omega\, e^{-i \omega t+i m \phi}{}_s\Slm(\theta){}_s\Slm^{*}(\theta') G_{\ell m}(r,r';\omega),  \label{Eq:TDGF}
\end{align}
\MC{Where did you get the above from? In Eq.2.35~\cite{Casals:Ottewill:2015} for Schwarzschild there's an extra $\Delta^s$ and $\Sigma$ (=$r^2$ in Schwarzschild)}
for some $c>0$,
where ${}_s\Slm$ are the spin-weighted spheroidal harmonics~\cite{Berti:2005gp,berti2006erratum}. 
Here we have made use of the axisymmetry and stationarity of Kerr space-time to set $t'=0$ and $\phi'=0$, without loss of generality.
The Fourier modes $G_{\ell m}$ of the Green function $\Gret$ are then themselves Green functions of the radial Teukolsky equation:
\begin{equation} \label{eq:radial teuk. eq.}
\left[\Delta^{-s }\frac{d}{dr}\left(\Delta^{s+1}\frac{d}{dr}\right)+\frac{K^2-2is (r-M)K}{\Delta}+4is \omega r-{}_{s}\lambda_{\ell m\omega}\right]
G_{\ell m}(r,r';\omega)=\delta(r-r')
\end{equation}
where $K\equiv (r^2+a^2)\omega-am$,
and ${}_{s}\lambda_{\ell m\omega}$ is an eigenvalue of the spin-weighted spheroidal harmonic equation. 

A radial Green function can be constructed from two linearly independent homogeneous solutions satisfying certain boundary conditions at infinity and at the horizon.
A physically-relevant pair of linearly independent solutions are the `ingoing' and `upgoing' solutions defined by the following boundary conditions:
\begin{align}
\Rin(r,\omega)&\sim \left\{\begin{array}{l l}
\Btrans\Delta^{-s}e^{-i \tilde{\omega}r_*}, & r\rightarrow r_+, \\
r^{-2s-1} \Bref e^{i \omega r_*} +r^{-1} \Binc e^{-i \omega r_*}, & r \rightarrow \infty.
\end{array}
\right. \\
\Rup(r,\omega)&\sim\left\{\begin{array}{l l}
C^{\text{inc}}e^{i \tilde{\omega}r_*}+C^{\text{ref}}\Delta^{-s}e^{-i \tilde{\omega}r_*}, & r\rightarrow r_+, \\
r^{-2s-1} \Ctrans e^{i \omega r_*}, & r \rightarrow \infty.
\end{array}
\right.
\end{align}
\MC{No subindices in the radial coefficients (or Wronskian)?}
where $\tilde{\omega}\equiv \omega-m\Omega_H$, $\Omega_H\equiv a/(r_+^2+a^2)$ is the angular velocity of the black hole
and $B^{\text{inc}/\text{ref}/\text{tra}}$ and  $C^{\text{inc}/\text{ref}/\text{tra}}$ are complex coefficients.
Here we have defined the tortoise coordinate $\tort$ via $\dfrac{d\tort}{dr}=\dfrac{(r^2+a^2)}{\Delta}$ as
\begin{equation}\label{eq:tortoise}
\tort=
r+\frac{2M}{r_+-r_-}\left\{r_+\ln\left|\frac{r-r_+}{2M}\right|-r_-\ln\left|\frac{r-r_-}{2M}\right|\right\}.
\end{equation}

It is convenient to define new solutions $\Rinhat$ and $\Ruphat$ 
which are `ingoing' and `upgoing' with transmission coefficient equal to one:
 \begin{align}\label{eq:hatted slns}
 \Rinhat\equiv \frac{\Rin}{\Btrans}, \quad \Ruphat\equiv \frac{\Rup}{\Ctrans}.
 \end{align}

Consequently, they satisfy the following boundary conditions:
\begin{align} \label{eq:f,near hor}
&
\Rinhat(r,\omega) \sim
\Delta^{-s}e^{-i\omega_+ \tort}
, \quad
\tort\to -\infty, 
\\& \Ruphat(r,\omega) \sim  r^{-1-2s}e^{+i\omega \tort}, \quad \tort\to +\infty \nonumber.
\end{align}

The boundary conditions (\ref{eq:f,near hor}) determine the two solutions $\Rin$ and $\Rup$ to the radial equation  uniquely for $\omega\in \mathbb{R}$.
These boundary conditions 
also define $\Rinup$ unambiguously for $\text{Im}(\omega)\ge 0$ when $\tort\in\mathbb{R}$.
In $\text{Im}(\omega) < 0$, with $\tort\in\mathbb{R}$, the solution $\Rup$ is defined by analytic continuation.

The radial Green function which specifically yields the {\it retarded} Green function via Eq.(\ref{Eq:TDGF}) can be expressed as
\begin{align}
\Glm(r,r';\omega)=-\frac{\Rinhat(r_<,\omega) \Ruphat(r_>,\omega)}{W
}, \label{eq:rGF}
\end{align}
where $r_<\equiv\min(r,r'), r_>\equiv\max(r,r')$, and $W$ is the constant Wronskian
\begin{align}\label{eq:Wronsk}
W
\equiv
\Delta^{s+1}\bar{W}(\Rinhat,\Ruphat)\equiv\Delta^{s+1}\left(\Rinhat\frac{d \Ruphat}{dr}-\Ruphat\frac{d\Rinhat}{dr}\right).
\end{align}
In the next section we give analytical MST expressions for the radial solutions and radial coefficients.


\section{MST Method}\label{eq:MST}

Many of the results presented will be given using the terminology of Mano, Suzuki and Takasugi \cite{Mano:Suzuki:Takasugi:1996}. For
ease of reading we will now give the relevant expressions for our calculations. For a complete disposition on the MST 
methodology we direct the reader to the review by Sasaki and Tagoshi~\cite{Sasaki:2003xr}.

The solutions satisfying the retarded boundary conditions of ingoing radiation at the horizon and upgoing at
infinity are given by MST as infinite sums of hypergeometric functions and irregular confluent hypergeometric functions in various forms depending on required radii of convergence. 
We note that these MST series yield a specific normalization  for the `in' and `up' solutions, which we shall give explicitly.

The horizon solution is given as a series of hypergeometric functions as
\begin{align}
\Rin&=e^{i \epsilon \kappa x}(-x)^{-s-i(\epsilon+\tau)/2}(1-x)^{i(\epsilon-\tau)/2}p_{\text{in}}^\nu(x), \nn \\
p_{\text{in}}^\nu&\equiv\sum_{n=-\infty}^{\infty}\an{n} p_{n+\nu}(x), \nn\\
p_{n+\nu}(x)&\equiv{}_2F_1(n+\nu+1-i\tau,-n-\nu-i\tau;1-s-i\epsilon-i\tau;x), \label{Eq:Rin 2F1}
\end{align}
where $x\equiv\omega (r_{+}-r)/(\epsilon\kappa)$, $\epsilon\equiv2 M \omega$, $\kappa\equiv\sqrt{1-q^2}$, $q\equiv a/M$ and $\tau\equiv(\epsilon-m q)/\kappa$. 
Here, the series coefficients $a_n$ are calculated using a three-term recurrence relation given by:
\begin{equation}
\alpha_n^\nu \an{n+1}+\beta_n^\nu \an{n}+\gamma_n^\nu \an{n-1}=0,
\label{Eq:anrecursion}
\end{equation}
where
\begin{align}
\alpha_n^\nu&=\frac{i\epsilon\kappa(n+\nu+1+s+i\epsilon)(n+\nu+1+s-i\epsilon)(n+\nu+1+i\tau)}{(n+\nu+1)(2 n+2 \nu+3)} ,\label{Eq:alpha}\\
\beta_n^\nu&=-\lambda-s(s+1)+(n+\nu)(n+\nu+1)+\epsilon^2+\epsilon(\epsilon-m q)+\frac{\epsilon(\epsilon-m q)(s^2+\epsilon^2)}{(n+\nu)(n+\nu+1)},\label{Eq:beta} \\
\gamma_n^\nu&=-\frac{i\epsilon\kappa(n+\nu-s+i\epsilon)(n+\nu-s-i\epsilon)(n+\nu-i\tau)}{(n+\nu)(2 n+2 \nu-1)}. \label{Eq:gamma}
\end{align} 
The parameter $\nu$ is then calculated to guarantee that $\an{n}$ is the minimal solution of Eq.(\ref{Eq:anrecursion}) both as $n\to \infty$ and as $n\to -\infty$.
For small $\epsilon$, this value admits the expansion
\begin{equation}
\nu=\ell+\nu_2 \epsilon^2+O(\epsilon^3),
\end{equation}
where
\begin{equation} \label{eq:nu2}
\nu_2\equiv \frac{1}{2\ell+1}\left(-2-\frac{s^2}{\ell(\ell+1)}+
\frac{\left(\left(\ell+1\right)^2-s^2\right)^2}{(2\ell+1)(2\ell+2)(2\ell+3)}-
\frac{\left(\ell^2-s^2\right)^2}{(2\ell-1)2\ell (2\ell+1)}\right).
\end{equation}
In App.\ref{sec:MSTexpansions} we give an expansions for $\nu$ up to order $\eps^4$ for spin-$0$.
In its turn, the value of $\nu$ chosen as indicated guarantees that the series in
 Eq.(\ref{Eq:Rin 2F1}) converges for all $|x|<\infty$. 

We note that the radial Teukolsky equation is invariant under complex conjugation together with $m\to -m$
and $\omega\to -\omega$. We shall therefore assume $\text{Re}(\omega)>0$ from now on without loss of generality.
A solution for $\Rup$ as a series of confluent hypergeometric functions is 
\MC{From what they say above Eq.139\cite{Sasaki:2003xr}, is the following just valid for $\text{Re}(\omega)>0$ (or $\omega>0$)?}
\begin{align}
	\Rup&= 2^{\nu}e^{-\pi \epsilon}e^{-i\pi(\nu+1+s)}
	e^{i\hat{z}}\hat{z}^{\nu+i\epsilon_+}(\hat{z}-\epsilon\kappa)^{-s-i\epsilon_+}
	\nonumber\\
	&\times\sum_{n=-\infty}^{\infty}i^n
	\frac{(\nu+1+s-i\epsilon)_n}{(\nu+1-s+i\epsilon)_n}
	\an{n}(2\hat{z})^n 
	U(n+\nu+1+s-i\epsilon,2n+2\nu+2;-2i\hat{z}), \label{eq:Rup series U}
\end{align}
where the $\an{n}$ series coefficients are the same as those in Eq.(\ref{Eq:Rin 2F1}),
 $\hat{z}\equiv \omega (r-r_-)=\epsilon\kappa (1-x)$ and $\epsilon_+\equiv (\epsilon+\tau)/2$.
We use $(z)_n$ to denote the Pochhammer symbol  $(z)_n=\Gamma(z+n)/\Gamma{z}$.
The series in Eq.(\ref{eq:Rup series U})  is convergent for $r>r_+$ when $\nu$ is calculated as mentioned above. 
In determining $\Rup$, MST also give another solution,  $R_+^\nu$, to the Teukolsky equation which has boundary condition
\begin{equation}
R_+^\nu\sim R_+^\text{tra}\frac{ e^{-i \omega \tort}}{r},\quad r\to \infty,
\end{equation}
where $R_+^\text{tra}$ is a coefficient that we determine below.
This solution can also be expressed as a series of confluent hypergeometric functions:
\begin{align}
R_{+}^{\nu}&= 2^{\nu}e^{-\pi \epsilon}e^{i\pi(\nu+1-s)}
\frac{\Gamma(\nu+1-s+i\epsilon)}{\Gamma(\nu+1+s-i\epsilon)}
e^{-i\hat{z}}\hat{z}^{\nu+i\epsilon_+}
(\hat{z}-\epsilon\kappa)^{-s-i\epsilon_+}\nonumber\\
&\times  \sum_{n=-\infty}^{\infty}i^n
\an{n}(2\hat{z})^n
U(n+\nu+1-s+i\epsilon,2n+2\nu+2;2i\hat{z}).
\label{eq:R+MST}
\end{align}
We shall use the solution in Eq.(\ref{eq:R+MST}) later. 
Finally, MST give expressions for $\Rin$ and $\Rup$ which are valid at infinity and the horizon respectively, however 
we will omit these here. Using all of these expressions, the asymptotic amplitudes can be calculated \cite{Sasaki:2003xr}, we give those relevant to our results:
\begin{align}\label{eq:Binc/ref/tra}
\Btrans=&\left(\frac{\epsilon \kappa}{\omega}\right)^{2 s}e^{i \kappa \epsilon_+(1+2\log\kappa/(1+\kappa))}\sum_{n=-\infty}^{\infty} \an{n}\nn,\\
B^{\rm inc}
=&\omega^{-1}\left[{K}_{\nu}-
ie^{-i\pi\nu} \frac{\sin \pi(\nu-s+i\epsilon)}
{\sin \pi(\nu+s-i\epsilon)}
{K}_{-\nu-1}\right]A_{+}^{\nu} e^{-i(\epsilon\ln\epsilon -\frac{1-\kappa}{2}\epsilon)},
\nn\\
B^{\rm ref}
=&\omega^{-1-2s}\left[{K}_{\nu}
+ie^{i\pi\nu} {K}_{-\nu-1}\right]A_{-}^{\nu}
e^{i(\epsilon\ln\epsilon -\frac{1-\kappa}{2}\epsilon)}, \nn\\
\end{align}
and\footnote{Note that there is typo in Eqs.155 and156~\cite{Sasaki:2003xr}, as the $z$ in the exponentials should be a $\hat{z}$.\MC{Check that no other $z$ should be a $\hat z$}}
\begin{align}
C^{\rm tra}
=&\omega^{-1-2s}e^{i(\epsilon\ln\epsilon -\frac{1-\kappa}{2}\epsilon)}A_{-}^\nu, \nn\\
R_+^{\rm tra}=&\omega^{-1}e^{-i(\epsilon\ln\epsilon-\frac{1-\kappa}{2}\epsilon)}A_{+}^\nu	.
\end{align}
Here, much complication is stored in the quantities $A_{+}^\nu,A_{-}^\nu$ and $K_\nu$. These are given by
\begin{align}\label{eq:A+/-}
&A_{+}^\nu=e^{-{\pi\over 2}\epsilon}e^{{\pi\over 2}i(\nu+1-s)}
2^{-1+s-i\epsilon}{\Gamma(\nu+1-s+i\epsilon)\over 
\Gamma(\nu+1+s-i\epsilon)}\sum_{n=-\infty}^{+\infty}\an{n},\\
&A_{-}^\nu=2^{-1-s+i\epsilon}e^{-{\pi\over 2}i(\nu+1+s)}e^{-{\pi\over 2}\epsilon}
\sum_{n=-\infty}^{+\infty}(-1)^n{(\nu+1+s-i\epsilon)_n\over 
(\nu+1-s+i\epsilon)_n}\an{n}. 
\nonumber
\end{align}
and
\begin{align}
K_{\nu}=&	\frac{e^{i\epsilon\kappa}(2\epsilon \kappa )^{s-\nu-r}2^{-s}i^{r}
	\Gamma(1-s-2i\epsilon_+)\Gamma(r+2\nu+2)}
	{\Gamma(r+\nu+1-s+i\epsilon)
	\Gamma(r+\nu+1+i\tau)\Gamma(r+\nu+1+s+i\epsilon)}
	\nonumber\\
	&\times \left ( \sum_{n=r}^{\infty}
	(-1)^n\, \frac{\Gamma(n+r+2\nu+1)}{(n-r)!}
	\frac{\Gamma(n+\nu+1+s+i\epsilon)}{\Gamma(n+\nu+1-s-i\epsilon)}
	\frac{\Gamma(n+\nu+1+i\tau)}{\Gamma(n+\nu+1-i\tau)}
	\,\an{n}\right)
	\nonumber\\
	&\times \left(\sum_{n=-\infty}^{r}
	\frac{(-1)^n}{(r-n)!
	(r+2\nu+2)_n}\frac{(\nu+1+s-i\epsilon)_n}{(\nu+1-s+i\epsilon)_n}
	\an{n}\right)^{-1},
	\label{eq:Knu}
\end{align}
where in this instance $r$ is an arbitrary integer chosen for convenience\footnote{We use the same symbol, $r$, for this integer number as for the radial coordinate so as to follow the same
notation as~\cite{Sasaki:2003xr}. It should be clear from the context when $r$ refers to  the radial coordinate and when to this arbitrary integer.} .

\section{Green function in the complex frequency domain}\label{sec:GF compl}


The Green function given by \eqref{Eq:TDGF} in its current form would require the homogeneous `in' and `up'
solutions for all $\omega\in\mathbb{R}$ for each $\ell$ and $m$ mode. 
Analytical progress can be made by means of a contour deformation on the complex-frequency plane of the real-frequency Fourier integral. 
Leaver gave significant insight into the
Green function in Schwarzschild spacetime by such deformation \cite{Leaver:1986}.
When deforming the contour, one must take into account the singularities of the Fourier modes of the Green function as dictated by 
Cauchy's theorem.
In Schwarzschild space-time, the singularities of the Fourier modes are: simple poles (quasi-normal modes) and a branch point at the origin
with a corresponding branch cut typically taken down the negative imaginary axis.
The contour deformation then means that the Green function may be obtained by the sum of the following contributions:
(1) a sum over the residues at the poles,
(2) an integral along a high-frequency arc and 
(3) an integral around a branch cut along the negative frequency axis. 
In his work, Leaver identified the low frequency portion of the branch cut integral with the late time behaviour of the Green function.  

\begin{figure}[htb!]
\centering
\includegraphics[width=.7\linewidth]{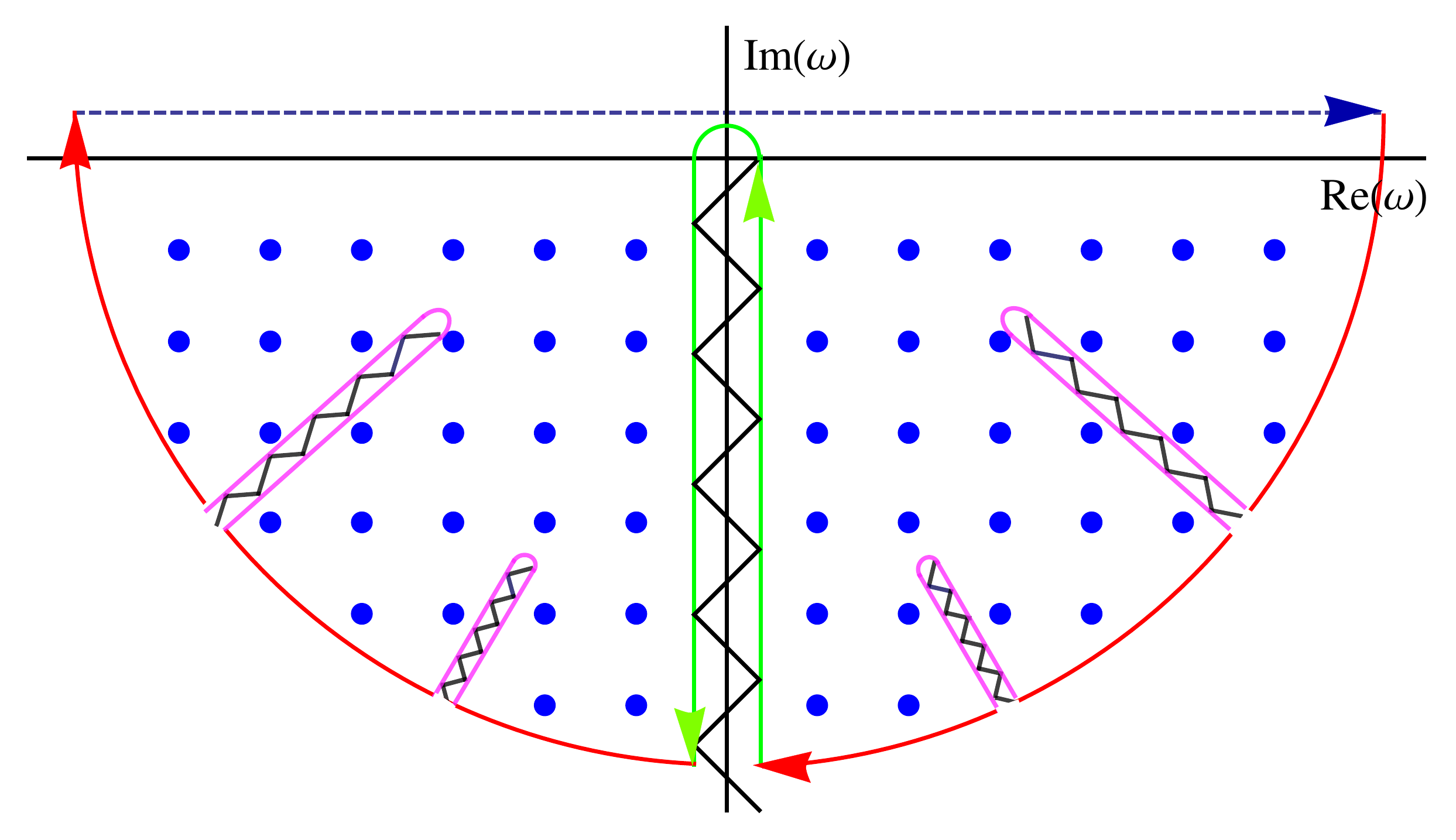}
\caption{
Contour deformation on the complex-frequency plane for the Green function of the Teukolsky equation in Kerr spacetime and
schematic representation of the singularities of its Fourier modes.
Dashed dark blue line: original Fourier-integration just above the real axis.
Red semi-circle: high-frequency arc integration.
Blue dots: simple poles of the Fourier modes  (quasi-normal modes).
Criss-crossed black lines and envolving curves: branch cuts of the Fourier modes and corresponding integration contours around them;
off the origin (with envolving green contour) for the branch cut of the radial functions;  off points away from the origin (with envolving light blue contours) for the branch
cuts of the angular functions.
Unlike the other singularities, these latter angular branch cuts and envolving integration contours are not present in Schwarzschild~\cite{Leaver:1986,Casals:2013mpa}.
We show in App.~\ref{sec:AngularCuts}, however that these angular cuts do not contribute to the full Green function.
[We note that in this figure we are ignoring any other extra cuts which the coefficients $\an{n}$ and/or the pararameter $\nu$ might have]
}
\label{fig:Kerrcontour}
\end{figure}

In Kerr spacetime, while the significant features of the poles giving QNMs and a branch cut leading to a late time tail remain, one also must account for branch points -- away from the origin -- in the spheroidal functions \cite{Oguchi70,BONGK:2004}.
In App. \ref{sec:AngularCuts} we show that these angular branch cuts are, however, spurious artefacts
of the spheroidal decomposition, which will vanish when we do the infinite sum over $\ell$ to obtain the full Green function. 
We show a schematic representation of the contour deformation and singularities in the complex-frequency plane
  in Fig.\ref{fig:Kerrcontour} (where, in the case of Schwarzschild, the angular branch cuts
are not present).

In this paper we are interested in the late-time behaviour of the Green function.
From asymptotic theory of Laplace transforms~\cite{Doetsch1974}, this  late-time behaviour will be dictated, after performing a Laplace transform on the Green function, 
by the uppermost singularities in the complex frequency plane of the Green function modes.
In the next section we argue that the only `physical' branch point in subextremal Kerr space-time is that at the origin, $\omega=0$.
Furthermore, Whiting~\cite{whiting1989mode} showed that there are no exponentially growing modes (i.e., with positive imaginary frequency) in subextremal Kerr.
Finally, on energy-conservation grounds, no quasi-normal modes may have real and non-superradiant frequency (i.e., $\omega\tilde{\omega}\ge 0$ and $\omega\neq 0$)~\cite{casals2016horizon}.
We therefore expect that the late-time behaviour of the Green function is given by the Green function modes near $\omega=0$.\MC{what about QNMs with super-radiant freq.?}


Our strategy for calculating the late-time behaviour of the Green function will be as follows: 

\begin{enumerate}
	\item express the discontinuity across the  branch cut  in the frequency domain radial Green function in terms of the discontinuity of the upgoing homogeneous solution;
	\item find an analytic expression for the discontinuity in the upgoing homogeneous solution in terms of known MST quantities;
	\item use this to calculate the induced discontinuity in the Wronskian;
	\item explicitly compute a low-frequency expansion of the branch cut contribution to the Green function;
	\item integrate the above branch cut expansion to obtain the behaviour of the Green function at late times.
\end{enumerate}

In Sec.\ref{sec:BC} we  deal with the above point 1  (see, specifically, Eq.(\ref{Eq:GBC})), point 2 (see Eqs.(\ref{eq:disc Ruphat}), (\ref{Eq:q eqtn})) and
point 3 (see Eqs.(\ref{Eq:Wp})--(\ref{Eq:WpWm})).
In Sec.\ref{sec:low freq r-indep} we deal with point  4 for the radius-independent part and in Sec.\ref{sec:low freq r-dep}
for the radius-dependent part.
Finally, in Sec.\ref{sec:late-time} we will deal with point 5 above.


\section{Branch cut}\label{sec:BC}

By a simple rescaling of the dependent variable, the radial Eq. (\ref{eq:radial teuk. eq.}) can be rewritten in Schr\"odinger-like form
(see Eqs.2.2 and 2.13~\cite{Hartle:Wilkins:1974}\footnote{We note that the independent variable in~\cite{Hartle:Wilkins:1974} is slightly different from 
the standard tortoise coordinate, but this should not affect the following conclusions about branch points.}).
It is easy to show that the potential in this Schr\"odinger-like equation goes like a constant term plus a term exponentially decreasing with the independent variable as
the horizon is approached. This means that, following the heuristic arguments in~\cite{Ching:1995tj}, 
the radial solution $\Rin$ is not expected to have a branch point at
$\tfreq=0$\MC{Given Eq.2.15~\cite{Hartle:Wilkins:1974}, why is it not $\Omega$ instead of $k$ in Eq.2.4~\cite{Hartle:Wilkins:1974}?}
On the other hand, the potential, after excluding the centrifugal barrier, goes like $\omega^2$ plus a term that decays slower than exponentially as radial infinity is approached.
This means that $\Rup$ is expected~\cite{Ching:1995tj} to have a branch point at the origin of the complex-frequency plane ($\omega=0$).

The MST series for the `in' and `up' radial solutions confirm the above expectations.
We first deal with the `in' solutions.
The representation in Eq.(\ref{Eq:Rin 2F1}) for `in' is in terms of hypergeometric functions which manifestly have no branch point in the complex-frequency plane, given
their analyticity properties as functions of their first three arguments~\cite{NIST:DLMF}.
The series coefficients $\an{n}$ and the renormalized angular momentum $\nu$, which appear in Eq.(\ref{Eq:Rin 2F1}), are also functions of $\omega$.
We show later on, however, that, at least to the order in $\epsilon$ to which we calculate them, these quantities possess no discontinuity along the negative imaginary axis.
In fact, we believe that these quantities have no discontinuity anywhere along the negative imaginary axis, as also expected by Leaver in~\cite{Leaver:1986a}, or, at least,
that if they happen to have any discontinuities along the negative imaginary axis, then these
do not contribute to the full Green function (e.g., they posses angular branch points, which, not only are away from the origin but also, as we show
in App.~\ref{sec:AngularCuts}, they do not contribute to the full Green function).
Alternatively, one could use the Jaff\'e series representation~\cite{Leaver:1986a} of the `in' solutions (see also Eq.73~\cite{Leaver:1986a} for the `up' solutions), 
which does not depend on $\nu$, although it has different series
coeffiicients whose analytic properties should be investigated.
\MC{Do you guys have a better way to phrase this or prove/justify that $\an{n}$ and $\nu$ have no BC down the NIA?
Are Eqs.(\ref{Eq:R BC}) and (\ref{Eq:GBC}) still valid all the way down the NIA even if $\an{n}$ and $\nu$ have BCs in other places (such as angular BCs)?}
For the `up' radial solution,  let us  consider its series representation Eq.(\ref{eq:Rup series U})  in terms of the 
irregular confluent hypergeometric functions $U(a,b,z)$.
These special functions contain a branch point at the origin of their third argument~\cite{NIST:DLMF}.
In our case, this means that the `up' solutions possess a branch point at $\omega=0$, as expected.
As is standard, we shall take the branch cut from $\omega=0$ to lie down the negative imaginary axis of the complex-$\omega$ plane.
In this paper we calculate the Green function modes along this branch cut. 
The following analytic continuation property~\citep{bk:AS} will be most useful:
\begin{align}
U(a,b,ze^{2 \pi i n})=(1-e^{-2 \pi i b n})\frac{\Gamma(1-b)}{\Gamma(1+a-b)}M(a,b,z)+e^{-2 \pi i b n}U(a,b,z),\quad n\in\mathbb{Z}^+, \label{Eq:PhiBC}
\end{align}
where $M(a,b,z)$ is the regular confluent hypergeometric function. 
We will use \eqref{Eq:PhiBC} to obtain an expression for the discontinuity in $\Ruphat$ across the negative imaginary frequency axis --  see Eq.(\ref{eq:delta Rup}) below.

We first establish some useful notation for the analysis on the branch cut.
We define an auxiliary variable $\fNIA\equiv i\omega$ to parameterise the frequency along the negative imaginary axis (it is $\fNIA>0$ along the cut). Henceforth,  $+/-$ {\it super}scripts denote functions evaluated respectively on the right/left of the branch cut, e.g.,
 $\Ruppm(r,\fNIA)\equiv\lim_{\rho\rightarrow 0^+}\Rup(r,-i\fNIA\pm\rho)$ with $\fNIA>0$.
 Also, the symbol `$\Disc$' will denote the difference between these two limits of a certain function, e.g.,
 $\Disc\Rup(r,\fNIA) \equiv \Rupp(r,\fNIA)-\Rupm(r,\fNIA)$.

The
asymptotic behaviour at radial infinity
of the `up' radial solution is the same in the limit of the frequency approaching the negative imaginary axis from the third or the fourth quadrant:
\begin{align}
\Ruppmhat
\sim r^{-1-2s}e^{\sigma \tort} ,\quad r\to\infty.
\end{align}
This implies that the difference between the two must be subdominant at infinity:
\begin{align}
\Disc\Ruphat\sim r^{-1}e^{-\sigma \tort},\quad r\to \infty. 
\end{align}
Therefore, this difference must be proportional to the (normalised) linearly independent solution $\hat{R}_+^\nu$:
\begin{align}
\Disc\Ruphat=i q(\sigma) \hat{R}_+^\nu, \label{Eq:R BC}
\end{align}
where the constant of proportionality is a `branch cut strength' function, $q(\sigma)$. 
Note that $q(\sigma)$ is a real-valued function in
Schwarzschild spacetime~\cite{Casals:Ottewill:2015,Leung:2003ix}
but
 in Kerr spacetime we have no reason to expect this. \\

The branch cut down the negative imaginary axis that $\Rup$ possesses is inherited by the 
Wronskian $W$ via Eq.(\ref{eq:Wronsk}) and by the 
Green function Fourier modes $G_{\ell m}$ 
via Eq.(\ref{eq:rGF}).
We now calculate the discontinuity in the  Green function modes across the cut from Eq.(\ref{eq:rGF}):
\begin{align}
\Disc \Glm(r,r';\fNIA)&=-\Rinhat(r_<,-i\fNIA)\left(\frac{\Rupphat(r_>)}{\Wbp}-\frac{\Rupmhat(r_>)}{\Wbm}\right) \nn\\
&=-\frac{\Rinhat(r,-i\fNIA) \Rinhat(r',-i\fNIA)}{\Wbp \Wbm} 
\Delta^{s+1}
\bar W(\Rupphat,\Rupmhat) \nn\\
&=-2 i \sigma \frac{q(\sigma)}{\Wbp\Wbm} \Rinhat(r,-i\fNIA) \Rinhat(r',-i\fNIA), \quad \fNIA>0,
 \label{Eq:GBC}
\end{align}
\MC{Check whether the $W(\Rupphat,\Rupmhat)$ above is ok with $\Delta^{s+1} W(\Rupphat,\Rupmhat)$}
where we have used Eq.(\ref{Eq:R BC}) and the fact that $\Rin$ possesses no branch cut. 
In the following two subsections, we shall obtain expressions for the Wronskian and for the branch cut strength $q(\sigma)$.

The contribution from the branch cut to the Green function $G(x,x')$ in Eq.(\ref{Eq:TDGF}) is then given by
\begin{equation}\label{eq:G_BC}
G_{BC}=
\sum_{\ell=|s|}^{\infty}\Disc G_{\ell},
\end{equation}
where we have defined
\begin{equation}\label{eq:Disc Gell}
\Disc G_{\ell}\equiv -2i \sum_{m=-\ell}^{\ell}e^{ i m \phi} \int_{0}^{\infty}d\fNIA\, e^{-\fNIA t}{}_s\Slm(\theta){}_s\Slm^{*}(\theta') \Disc \Glm(r,r';\fNIA),
\end{equation}
and where the spin-weighted spheroidal harmonics are meant to be evaluated at $\omega=-i\fNIA$.


\subsection{Wronskian discontinuity}

Our expression for the discontinuity of the Green function Eq.(\ref{Eq:GBC}) involves calculating the Wronskian of the two homogeneous solutions on either side of the negative imaginary axis, and taking their product: $\Wbp\Wbm$. We will now express this in terms of known asymptotic amplitudes evaluated entirely on the right hand side of the branch cut, i.e., on the 4th quadrant.
Straight-forwardly, we have
\begin{align} \label{Eq:Wp}
\Wbp=W(\Rinhat,\Rupphat)&=2 \sigma \frac{\Bincp}{\Btransp}.
\end{align}
We wish to write $W^-$ in terms of functions evaluated on the right side of the cut also. Using Eq.(\ref{Eq:R BC}),
\begin{align} \label{Eq:Wm}
\Wbm= W(\Rinhat,\Rupmhat)&= W(\Rinhat,\Rupphat)-i q(\sigma) W(\Rinhat,\hat{R}^\nu_+) \nn\\
&=2 \sigma \frac{\Bincp}{\Btransp}-2 i \sigma q(\sigma) \frac{\Brefp}{\Btransp},
\end{align}
so that 
\begin{align}
\Wbp \Wbm=\left(2 \sigma \frac{\Bincp}{\Btransp}\right)^2-4 i \sigma^2 q(\sigma) \frac{\Bincp \Brefp}{(\Btransp )^2}. \label{Eq:WpWm}
\end{align}
Here, 
$\Bincp$, $\Brefp$ and $ \Btransp$
 must be evaluated by analytically continuing 
their MST expressions from the real axis down to the right of the negative imaginary axis. Practically, this amounts to setting $\epsilon=2 M \sigma e^{-i \pi/2}$ with $\sigma>0$
 in our formulas. We will then expand in small $\sigma$.


\subsection{Branch cut strength}
  
We now derive an expression for the branch cut strength function $q(\sigma)$. 
We begin by writing Eq.(\ref{eq:Rup series U}) for $\Ctrans\Ruphat$ succinctly
as
\begin{align}
\Ruphat(r,\omega)=\frac{1}{\Ctrans} f(\epsilon)\sum_{n=-\infty}^{\infty} A_n(\epsilon)U(a,b,-2 i \hat{z}),
\end{align}  
where
\begin{align}\label{eq:Rup vars}
	f(\epsilon)&\equiv 2^{\nu}e^{-\pi \epsilon}e^{-i\pi(\nu+1+s)}
e^{i\hat{z}}\hat{z}^{\nu+i\epsilon_+}(\hat{z}-\epsilon\kappa)^{-s-i\epsilon_+},\\
	A_n(\epsilon)&\equiv i^n
\frac{(\nu+1+s-i\epsilon)_n}{(\nu+1-s+i\epsilon)_n}
\an{n}(2\hat{z})^n,\nonumber \\ 
	a&\equiv n+\nu+1+s-i\epsilon, \nonumber\\
	b&\equiv 2n+2\nu+2.\nonumber
\end{align}	
After an anticlockwise rotation of `$2 \pi$' in the complex-frequency plane we have (assuming $s\in\mathbb{Z}$):
\begin{align}
	\hat{z}&\to \hat{z} e^{2 \pi i},\nonumber \\
\Ctrans(\epsilon e^{2 \pi i})&=\Ctrans(\epsilon)e^{-2 \pi \epsilon},\nonumber \\
f(\epsilon e^{2 \pi i})&=f(\epsilon) e^{2 \pi i \nu},\nonumber \\
A_n(\epsilon e^{2 \pi i})&=A_n (\epsilon).\nonumber
\end{align}  
Making use of these together with Eq.(\ref{Eq:PhiBC}) with $n=1$ and the 
 identity~\citep{bk:AS}
\begin{align}
M(a,b,-2 i \hat{z})=\frac{\Gamma(b)}{\Gamma(b-a)}e^{-a \pi i}U(a,b,-2 i \hat{z})+\frac{\Gamma(b)}{\Gamma(a)}e^{(b-a)i \pi}
e^{-2 i \hat{z}}U(b-a,b,2 i \hat{z}),
\end{align}
we find that\MC{I think in the eq. below there was a wrong overall minus sign. So I've changed it and carried the change all the way to Eq.(\ref{Eq:q eqtn}),
which now agrees, for $a=0$, with Eqs.5.10 and 5.11~\cite{Casals:Ottewill:2015} in Schwarzschild. I have also implemented this sign change to all the expansions for
$q$ below. My concern, though, is that you had seen agreement with the numerics
for the Green function with your (I believe, wrong) sign? so, for the moment, I haven't changed the sign in any of the expansions below for $\Disc G$}
\begin{align}\label{eq:delta Rup}
\Disc\Ruphat= & -\frac{f(\epsilon)}{\Ctrans} \sum_{n=-\infty}^{\infty}  A_n(\epsilon) \\
\times &\left[\left(e^{2 \pi i \nu}e^{2 \pi \epsilon}\left(1-e^{-2 \pi i b}\right) 
\frac{\Gamma(1-b)}{\Gamma(1+a-b)} \frac{\Gamma(b)}{\Gamma(b-a)}e^{-a \pi i} \right.  +e^{-2 \pi i \nu}e^{2 \pi \epsilon}-1\right)U(a,b,-2 i \hat{z}) \\
&\left.+e^{2 \pi i \nu}e^{2 \pi \epsilon}\left(1-e^{-2 \pi i b}\right) 
\frac{\Gamma(1-b)}{\Gamma(1+a-b)}\frac{\Gamma(b)}{\Gamma(a)}e^{(b-a)i \pi}e^{-2 i \hat{z}}U(b-a,b,2 i \hat{z}) \right] .
\end{align}
We focus first on the coefficient of $U(a,b,-2 i \hat{z})$. Using properties of the $\Gamma$-function and putting in $a$ and $b$ 
explicitly, this coefficient is equal to
\begin{align}
&e^{2 \pi i \nu}e^{2 \pi \epsilon}\left(1-e^{-4 \pi i \nu}\right)\frac{\sin(\pi(\nu+i \epsilon))}{\sin(2 \pi \nu)}e^{-i \pi( \nu-i \epsilon)}
 +e^{-2 \pi i \nu}e^{2 \pi \epsilon}-1 \nonumber \\
 &=(1-e^{2 \pi \epsilon}e^{-2 \pi i \nu})+e^{-2 \pi i \nu}e^{2 \pi \epsilon}-1 
 =0.\nonumber
\end{align} 
This leaves us with the $U(b-a,b,2 i \hat{z})$ term, which we want to write in terms of $\hat{R}_+^\nu$.
As with $\Rup$, for brevity we write Eq.(\ref{eq:R+MST}) for  $R_+^\nu$ as
\begin{align}
R_+^\nu=g(\epsilon)\sum_{n=-\infty}^{\infty} B_n(\epsilon)U(b-a,b,2i \hat{z}),
\end{align}
where $a$ and $b$ are as in Eq.(\ref{eq:Rup vars}) above. With a little examination, we can note the relation
\begin{align}
g(\epsilon) B_n(\epsilon)=e^{2 \pi i \nu}e^{-2 i \hat{z}}\frac{\Gamma(b-a)}{\Gamma(a)}f(\epsilon)A_n(\epsilon),
\end{align}
from which we can write the remaining part of $\Disc\Ruphat$ as
\begin{align}\label{eq:dRuphat intermid}
\Disc\Ruphat&=-\frac{g(\epsilon)}{\Ctrans}\sum_{n=-\infty}^{\infty} B_n(\epsilon) e^{ \pi \epsilon}\left(1-e^{-4 \pi i \nu}\right) \frac{\Gamma(1-b)}{\Gamma(1-(b-a))}\frac{\Gamma(b)}{\Gamma(b-a)}e^{i \pi(n+\nu-1-s)}U(b-a,b,2 i \hat{z}).
\end{align}
The $\Gamma$-functions simplify as
\begin{align}
\frac{\Gamma(1-b)}{\Gamma(1-(b-a))}\frac{\Gamma(b)}{\Gamma(b-a)}&=\frac{\sin(\pi(b-a))}{\sin(\pi b)}\nonumber \\
&=\frac{\sin(\pi(n+\nu+1-s+i \epsilon))}{\sin(\pi (2n+2\nu+2))}\nonumber\\
&=\frac{\sin(\pi(\nu+i \epsilon))}{\sin(2\pi\nu )}e^{i \pi(n+1-s)}\nonumber\\
&=\frac{e^{\pi(\epsilon+i \nu)}(1-e^{-2 \pi (\epsilon-i \nu)})}{1-e^{4 i \pi \nu}}e^{i \pi(n+1-s)}.\nonumber
\end{align}
Plugging this into Eq.(\ref{eq:dRuphat intermid}), using Eq.(\ref{eq:Binc/ref/tra}) and simplifying, we arrive at 
\begin{align}\label{eq:disc Ruphat}
\Disc\Ruphat&=\frac{1}{\Ctrans}
\left(e^{2\pi(\epsilon-i\nu)}-1\right)
R_+^\nu  \\
&=\frac{R_+^\text{tra}}{\Ctrans}
\left(e^{2\pi(\epsilon-i\nu)}-1\right)
\hat{R}_+^\nu.\nonumber
\end{align}
Comparing with Eq.\eqref{Eq:R BC} we obtain an analytic expression for the branch cut strength function:
\begin{align}
q(\fNIA)=i\frac{A_+^\nu}{A_-^\nu}\omega^{2 s}\epsilon^{-2 i \epsilon}e^{i \epsilon(1- \kappa)}\left(1-e^{2\pi(\epsilon-i\nu)}\right). \label{Eq:q eqtn}
\end{align}
This expression is valid 
 for all integer values of $s$. All functions here are analytic  at $\epsilon=0$
 and so can be expanded in Taylor series about this point. 


\section{Low-frequency Expansion of the Radius-Independent Part of the Branch Cut Modes}\label{sec:low freq r-indep}

We wish to calculate a small-frequency expansion of the BC modes $\Disc \Glm$  in Eq.(\ref{Eq:GBC}).
In the next section we will calculate the expansion for the radial function; 
in App.\ref{sec:MSTexpansions} we will do it for the series coefficients $\an{n}$ and for $\nu$;
in this section we  do it for the other radius-independent quantities:
the BC strength as in Eq.(\ref{Eq:q eqtn}) and the Wronskian factor that appears in Eq.(\ref{Eq:WpWm}).
The Wronskian factor depends on $B^{inc/ref/tra}$, which will be calculated via Eq.(\ref{eq:Binc/ref/tra}). 
This equation involves 
$A_{\pm}^\nu$, obtainable from Eqs.(\ref{eq:A+/-}), and
$K_{\nu}$ and $K_{-\nu-1}$, obtainable from Eq.(\ref{eq:Knu}).
These latter quantities are the most time-consuming ones to calculate.
It proves useful to factor out $K_{\nu}=e^{K_{\nu}^{\ph}} \check{K}_{\nu}$ 
and $K_{-\nu-1}=e^{K_{-\nu-1}^{\ph}} \check{K}_{-\nu-1}$, 
where
\begin{align}
 K_{\nu}^{\ph}&=i \epsilon  \kappa -\nu  \log (2 \epsilon  \kappa ), \label{Eq:KnuPh}\\
K_{-\nu-1}^{\ph}& =i \epsilon  \kappa +\nu  \log (2 \epsilon  \kappa ),
 \end{align}
 and $ \check{K}_{\nu}$  and $\check{K}_{-\nu-1}$ turn out to be polynomials in $\sigma$
 \MC{I took this from $\ell=m=0$ but is this true $\forall \ell$?}.
 Throughout this section we will use a check symbol above a quantity 
 and a `ph' superscript to factor the quantity into its exponential terms as above.

Without further ado, we now proceed to give the small-frequency expansions of the various quantities obtained using the corresponding equations
just referred to. We will calculate the expansions for spin $s=0$, up to the first five leading orders (if one considers a $\log(\sigma)$ term
as yielding a different order) in a general $\ell>2$ expression and separately for the cases $\ell=m=0$ and $\ell=|m|=1$ (since these
cases are not reproduced
by the general $\ell>2$ expression).


\subsection{Mode $\ell=m=0$}


For the mode $\ell=m=0$ we obtain
\begin{align}
	\check{K}_{\nu}=&\frac{9}{7}\Bigg[1+\left(\frac{4}{3}+2 \gamma_E -\frac{\kappa }{3}\right) M\sigma +
	\bigg(2 \gamma_E ^2(\gamma_E-1) -\frac{4   \left(148 q^2-6739\right)}{2205}
	-\frac{\pi ^2}{3}+ 
	\bigg(\frac{ \pi ^2}{1-q^2}-\frac{2}{3}-\gamma_E \bigg)\frac{2 \kappa}{3}   \bigg) M^2\sigma ^2
	\nn \\   &
	+ \bigg(
\frac{4 \gamma_E }{3}  (  \gamma_E ^2-5)
   +\frac{4 \left(23169-2608 q^2\right)}{6615}
-\frac{2 \pi ^2}{3} 
+\gamma_E  \left(\frac{32   \left(1256-37 q^2\right)}{2205}-\frac{2 \pi ^2}{3}\right)-
\nn \\   &
\frac{4 \left(4-q^2\right)   \psi ^{(2)}(1)}{3 \left(1-q^2\right)}+\kappa  \bigg(\frac{47 q^2-6367}{1323}
   +\frac{\left(23-q^2\right) \pi ^2}{9 \left(1-q^2\right)}+\frac{2\gamma_E}{3}    \left(1-\gamma_E +\frac{2 \pi ^2}{1- q^2}\right)-   \frac{4 \psi   ^{(2)}(1)}{1-q^2}\bigg)\bigg)M^3\sigma ^3\Bigg]
   +o\left(\sigma^3\right),
\end{align}
where $\gamma_E$ is Euler's constant and $\psi^{(n)}(z)$ is the polygamma function, which is the $n$th-derivative of the digamma function $\psi(z)$.
Similarly,
\begin{align}
	\check{K}_{-\nu-1}&=\frac{2 i}{7}\Bigg[1+\left(\frac{2}{3}+\gamma_E + \kappa \right) 2M\sigma +
	\bigg(2 \gamma_E ^2\left(\gamma_E  +\frac{11}{3}+2 \kappa \right)+\frac{766 q^2-17613}{2205}-\frac{\pi ^2}{3}+\left(8+\frac{2 \pi ^2}{1-q^2}\right) \frac{\kappa}{3}
\bigg) M^2\sigma ^2
		 \nn \\  &	
	+ \bigg(4 \gamma_E ^2\left(\frac{ \gamma_E}{3}    +3+ \kappa \right)-\frac{40
	   \left(428+25 q^2\right)}{1323}+\frac{8 \pi	   ^2}{9}-\bigg(\frac{ 11094+101 q^2}{441}+\frac{2 \left(2-q^2\right) \pi ^2}{3 \left(1-q^2\right)}\bigg) \kappa
	      \nn \\   &	 
	+\frac{2\gamma_E}{3}  \bigg(\frac{
	   \left(766 q^2-10753\right)}{735}-\pi ^2
	   +\left(11+\frac{\pi	   ^2}{1- q^2}\right) 2\kappa \bigg)
	   -\left(\frac{ 4-q^2}{3 }+ \kappa \right)\frac{ 4\psi ^{(2)}(1)}{1-q^2}\bigg)M^3\sigma ^3\Bigg]   +o\left(\sigma^3\right).
\end{align}

Continuing this analysis to the $A_\pm^{\nu}$ terms, we obtain
\begin{align}
\check{A}_{+}^\nu=&\frac{7}{9}\Bigg[1+\left(\frac{\kappa }{3}-4 \gamma_E \right) M\sigma +\left(\frac{2 \left(183+74
   q^2\right)}{2205}+2 \gamma_E ^2-\frac{ \gamma_E  \kappa }{3}\right)2 M^2\sigma ^2
       \nn \\   &
   +
   \bigg(
8 \gamma_E ^2  \left(\kappa   -4 \gamma_E\right)
   -\frac{32 \left(183+74 q^2\right) \gamma_E }{735}
+\frac{28   \pi ^2}{3}+\frac{\left(6343+949 q^2\right) \kappa }{2205}+8 \psi ^{(2)}(1)\bigg)\frac{M^3\sigma ^3} {3}\Bigg]  +o\left(\sigma^3\right),	
\end{align}
\begin{align}
\check{A}_{-}^\nu =&\frac{7}{9}\Bigg[1-\left(4+5 \kappa \right) \frac{M\sigma}{3} +\left(\frac{4 \left(1591-587
   q^2\right)}{245}+20 \kappa \right)\frac{ M^2\sigma ^2}{9}+\bigg(4 \left(4364   q^2-2577\right)
+\left(1807 q^2-16651\right) \kappa\bigg)\frac{ M^3\sigma ^3}{6615}   \Bigg]  +o\left(\sigma^3\right),
 \end{align}
 and
\begin{align}
  A_{\pm}^{\nu,\ph}&=-\frac{\pi  \epsilon }{2}\pm\frac{i \pi  (1+\nu )}{2} -(1\pm i \epsilon ) \log (2).
\nonumber
\end{align}
While all these quantities together are sufficient for calculating the branch cut strength $q(\sigma)$, we still need the $\Binc,\Bref$ and $\Btrans$ coefficients
 which appear in the Wronskian expression. 
 \MC{Are all the  coeffs. below on the right of the cut? Yes. so  we should add a superscript $+$. but then there would be 3 supercripts (inc,+,ph).
Maybe write explicitly $B^{\text{ref},+}=\Bref_{\ph}\check{B}^{\text{ref}}$, etc}
The expansions for the coefficients that we  give, when evaluated on the cut, are meant to be by taking the limit from the fourth quadrant - we do not
include a superscript `+' to not overburden the notation.
 The expansions for these coefficients for $\ell=m=0$ are:
\begin{align}
\check{B}^{\text{inc}}&=\frac{14}{9\epsilon}\Bigg[1
-\left(\gamma_E +\frac{ \kappa }{3}\right) 2M\sigma +\bigg(\frac{29639+347
   q^2}{2205}+2 \gamma_E ^2+\frac{(8 i -\pi) \pi }{3}+\left(\frac{2 \pi ^2}{3(1-  q^2)}-4\right) \kappa \nn \\
   & +\frac{2\gamma_E}{3}  \left(2 \kappa-11\right)-\frac{8}{3}
   \log (4M \kappa \sigma )\bigg)M^2\sigma ^2 +
    \bigg(\frac{2}{3} \left(19-q^2\right)
   +\frac{2\gamma_E ^2}{3} \left(11- \kappa-\gamma_E \right)+8 i \pi +\frac{20 \pi ^2}{9} \nn \\
   &+\left(\frac{28 i \pi }{9}-\frac{2 \left(6801+268 q^2\right)}{6615}+\frac{
   \left(12-q^2\right) \pi ^2}{ \left(1-q^2\right)9}\right) \kappa \nn \\
   & +\frac{\gamma_E }{3} \left(\pi  (\pi-8 i )
   -\frac{ \left(47279+347 q^2\right)}{735}+\left(\frac{11}{3}-\frac{\pi ^2}{ 1-q^2}\right) 2\kappa
   \right) \nn \\
   &+\left(\frac{2 \gamma_E }{3}-\frac{7 \kappa }{9}-2\right)4 \log (4M \kappa \sigma   )-\left(\frac{ 2+q^2}{3 }+ \kappa
  \right) 2\psi ^{(2)}(1)\bigg)\frac{2M^3\sigma ^3}{1-q^2}	\Bigg]+o\left(\sigma^3\right), 
   \end{align}
   \MC{There's a $\epsilon$ in the denominator of the overall factor above? If so, maybe we shoould write it all in terms of either $\epsilon$ or $\sigma$.Also, I should then change the last $o$}
   \begin{align} \label{eq:check Btra ell=0}
   \check{B}^{\text{tra}}&=\frac{7}{9}\Bigg[1+\frac{\kappa  M\sigma }{3}+\frac{8 \left(183+74 q^2\right) M^2\sigma
   ^2}{2205}+\frac{\left(6343+949 q^2\right) \kappa  M^3\sigma ^3}{6615}\Bigg]+o\left(\sigma^3\right),
   \end{align}
   \begin{align}\label{eq:ph Binc/ref/tra ell=0}
   \check{B}^{\text{ref}}&=\frac{14}{9\epsilon}\Bigg[1+2\left( \gamma_E -\frac{4 \kappa }{3}\right) M\sigma + \bigg(\frac{45809-7003
   q^2}{2205}+2 \gamma_E \left(\gamma_E-\frac{11+8 \kappa }{3}\right)
+      \frac{\pi ^2}{3} \left( \frac{2  \kappa }{1- q^2}-1\right)
-\frac{8}{3} \log (4M \kappa \sigma   )\bigg)M^2\sigma ^2
      \nn \\   &
   + \bigg(
   \frac{2\gamma_E^2 }{3}\left(\gamma_E-11-4 \kappa \right) 
   -\frac{8 \pi ^2}{9}
   +\left(\frac{7243   q^2-98109}{735}+\frac{\left(15-4 q^2\right) \pi ^2}{1-q^2}\right)\frac{ \kappa}{735} +
       \nn \\   &   
 \frac{  \gamma_E }{3}
   \bigg(\frac{ 45809-7003 q^2}{735}-\pi   ^2
   +\left(\frac{26}{3}+\frac{\pi ^2}{1- q^2}\right) 2\kappa\bigg)-\left(2   \gamma_E +\frac{ \kappa }{3}\right) \frac{4}{3}\log (4M \kappa \sigma )
   -   \left(\frac{4  \left(4-q^2\right)}{3}+ \kappa \right) \frac{2\psi   ^{(2)}(1)}{1-q^2}\bigg)2M^3\sigma ^3\Bigg]
       \nn \\   &   
   +o\left(\sigma^3\right),
      \end{align}
       \MC{There's a $\epsilon$ in the denominator of the overall factor above? If so, maybe we shoould write it all in terms of either $\epsilon$ or $\sigma$.Also, I should then change the last $o$}
and
   \begin{align}
   \Binc_{\ph}&=\frac{i \pi }{2}-\log (2)+  \left(1+2 i \pi +\kappa -2 \log (4     \sigma )\right)M\sigma+\nu  (i \pi -\log (4M \kappa \sigma )),\nn\\
   \Btrans_{\ph}&=\left(1+\kappa  +2   \log (\kappa )\right)M\sigma,\nn \\
   \Bref_{\ph}&=-\frac{i \pi }{2}-\log (2)-\nu  \log (4M \kappa \sigma )- \left(1-3 \kappa-2 \log (4   \sigma )\right)\sigma.
\end{align}
With these we can proceed to the results for the Wronskian factor and the $q$ function.


While we find the branch cut strength $q$ to be  quite simple even with the exponential terms included, it is useful to show it both
with and without factoring out the exponential terms. This is so as to see how it will behave with $\ell$ \MC{But we're just doing it for $\ell=0$?}
 and to find further cancellations when evaluating Eq.(\ref{Eq:GBC}). We have, for $\ell=m=0$,
\begin{align}
	\check{q}(\sigma)&=-4\pi M\sigma\Bigg[1+\left(\frac{11}{3}-4 \gamma_E +2 \kappa \right)M \sigma +\left(3-q^2+
	2 \gamma_E \left(2\gamma_E	-\frac{11}{3}	+\frac{2 \pi (\pi+7 i ) }{3}-2 \kappa \right)+\frac{11 \kappa   }{3}\right)2 M^2\sigma ^2 \nn \\
   &+ \bigg(8\gamma_E \left(\gamma_E \left(\frac{11-4   \gamma_E}{3}+2 \kappa \right)+
   a^2-3+\frac{\pi(\pi+7 i ) }{3} 
   -\frac{11 \kappa   }{3}\right)
   -\frac{22\pi \left( \pi+7 i\right)}{9}-\frac{ 229+1062 q^2}{135}
   \nn \\   &
   +\left( 7-q^2-\pi  (\pi-7 i) \right) \frac{4\kappa}{3} +\frac{8 \psi   ^{(2)}(1)}{3}\bigg)M^3\sigma ^3\Bigg]  
   +o\left(\sigma^4\right), 
   \end{align}
   and
   \begin{align}
   q^{\ph}&=i \pi  (1+\nu )+2   (1-\kappa -2 \log (4 M \sigma ))M\sigma.
\end{align}
From this we see explicitly that the phase will not contribute any extra factors of $\sigma$, but there will be an alternating sign like $(-1)^{\ell+1}$\MC{How can one infer a general-$\ell$ behaviour from a $\ell=0$ expression?}. The explicit full form of $q$ for $\ell=0$ is
\begin{align}\label{eq:q ell=0}
q(\sigma)=&4 \pi M \sigma\left(1+ \left(\frac{17}{3}-4 \cdot\text{Eulerlog}\right) M\sigma+
\left(2 \left(6 \cdot\text{Eulerlog}^2-17 \cdot\text{Eulerlog}+10\right)-\pi ^2\right)\frac{2 M^2\sigma ^2}{3} \right. \nonumber \\
&+ \left(2\pi ^2\left(4 \cdot\text{Eulerlog}-\frac{17}{3}\right)-\frac{8 q^2}{5}-\frac{1}{45} \left(1440 \cdot\text{Eulerlog}^3-6120 \cdot\text{Eulerlog}^2+7200 \cdot\text{Eulerlog}-1031\right)
\right. \nonumber \\&\left.\left.
+8 \psi ^{(2)}(1)\right)\frac{M^3\sigma ^3}{3}\right)
+o\left(\sigma^4\right),
\end{align} 
where EulerLog$\equiv\log(4 \sigma)+\gamma_E$.


Since the Wronskian comes in two pieces one needs to check the common phase between the two to determine an overall exponential factor, which is quite a simple task given everything we have above. After combining some simplifying exponential terms involving `$\log(2)$' and `$i\pi$', this gives  
\begin{align}\label{eq:check W fac ell=0}
	\check{W}^+\check{W}^-&=1-(2\gamma_E+\kappa ) 2M\sigma +
	     \nn \\   &
	\bigg(\frac{245-17 q^2}{9}+8
   \gamma_E ^2-\frac{2 \pi (\pi+14 i) }{3}+\left(\frac{ \pi ^2}{3(1- q^2)}-2\right)
 4  \kappa 
 +4\gamma_E  \left(2 \kappa-\frac{11}{3} \right)-\frac{16}{3} \log (4M \kappa \sigma
   )\bigg)M^2\sigma ^2+
        \nn \\   &
    \bigg(\frac{8}{3} \left(23-5 q^2\right)-\frac{32 \gamma_E
   ^3}{3}+\frac{64 \pi ^2}{9} 
   +16\gamma_E ^2 \left(\frac{11}{3}- \kappa
   \right)+\left(\frac{ 5 q^2-197}{3}+28 i \pi +\frac{\left(14-3
   q^2\right)2 \pi ^2}{(1- q^2)3}\right) \frac{2\kappa}{3}
     \nn \\   &
     +\frac{\gamma_E }{3} \left(\frac{4}{3} \left(17   q^2-317\right)+112 i \pi +8 \pi ^2+\left(17-\frac{2 \pi ^2}{   1-q^2}\right) 8\kappa \right)
      \nn \\   &
   +\left(\frac{64 \gamma_E }{3}-\frac{16 \kappa   }{3}-32\right) \log (4M \kappa \sigma )-\left(\frac{2+q^2}{3}+ \kappa \right)\frac{8 \psi ^{(2)}(1)}{1-q^2}\bigg)M^3\sigma ^3
   +o\left(\sigma^3\right), 
   \end{align}
   and
   \begin{align}\label{eq:ph W fac ell=0}
   (W^+W^-)^{\ph}&=2\nu  ( i \pi - \log (4M \kappa \sigma ))-4 M \sigma  \log (4M \kappa \sigma ),
\end{align}
from which we can see an immediate contribution in the general-$\ell$ case from the exponential of $\sigma^{2\ell}$\MC{Again, how can one infer a general-$\ell$ behaviour from a $\ell=0$ expression?}.

Finally we present the ratio of the branch cut strength and the Wronskian product, the key quantity required in 
Eq.~(\ref{Eq:GBC}) to determine the discontinuity in the radial Green function. In this final step we will explicitly include much of the phase terms since they source the logarithmic contributions to the late time tail first reported in Schwarzschild in~\cite{Casals:Ottewill:2015}.
The ratio of the branch cut strength and the Wronskian product is
\begin{align}
	\frac{q(\sigma)}{W^+W^-}&=e^{(1-\kappa+2\log\kappa)M\sigma}4\pi M\sigma\bigg[1+M \left(\frac{11}{3}+4 \kappa \right) \sigma+\frac{1}{3} \left(44 \gamma +\frac{1}{3} \left(-119-73 q^2\right)-\frac{4 \pi ^2}{\kappa }+68 \kappa +44 \log (4M \kappa \sigma
   )\right)   M^2 \sigma ^2\nn \\
   &-\frac{1}{135} M^3 \sigma ^3 \bigg(11544+8767 q^2-  (11580+10080 \kappa )(\gamma_E+\log (4M \kappa \sigma ))+ \frac{120 \pi ^2 (11+17 \kappa )}{\kappa }+60 \left(119+25
   q^2\right) \kappa \nn \\
   &+\frac{2160}{\kappa ^2}\left(1+\kappa\right) \zeta (3)\bigg)
	\bigg]+o(\sigma)^4.
\end{align}


\subsection{Mode $\ell=|m|=1$\MC{But the results below are only valid for $m=1$, not for $m=-1$?}}

In Sec.\ref{sec:late-time} we will evaluate the Green function on the equator ($\theta=\pi/2$).
By the symmetries\MC{say which ones} of the Teukolsky equation, $\Disc G_{1,1}+\Disc G_{1,-1}=2 \text{Re}(\Disc G_{1,1})$ on the equator.
Therefore, on the equator, we  only need calculate $m=1$ quantities, not $m=-1$. 
All results in this subsection are for $\ell=m=1$.
Much of the details of this calculation are quite similar to the previous subsection for $\ell=0$.
Therefore, we only give the main results needed, that is the expansions for the transmission coefficient for the inner solution, the branch cut strength and the Wronskian. The transmission coefficient is
\begin{align}\label{eq:check Btra ell=m=1}
  \check{B}^{\text{tra}}&=	1- \left(\frac{25 i q}{361}-\kappa \right) M\sigma + \left(\frac{3416}{9025}-\frac{310997 q^2}{3258025}+\frac{236 i q \kappa }{1805}\right)
	 M^2  \sigma ^2+ \bigg(\frac{57390747 i q}{22806175}-\frac{840440462 i q^3}{1646605835}\nn \\
	   &-\frac{19669 \kappa }{9025}+\frac{1963303 q^2 \kappa
	   }{3258025}\bigg)M^3 \sigma ^3+o(M^3\sigma^3),
\end{align}	   
	   and
	   \begin{equation}\label{eq:ph Btra ell=m=1}
	   \Btrans_{\ph}=  (1+\kappa +2 \log (\kappa ))M \sigma-\frac{i q (1+\kappa +2 \log (\kappa ))}{2 (1+\kappa )}.
\end{equation}

We find the branch cut strength to be, again both with and without factoring out the exponential terms,\MC{Are all the following for $m=1$? It'd be good to also give $m=-1$ so that people can do calculations off the equator}
\begin{align}
\check{q}(\sigma)&=-4\pi M \sigma\Bigg[1+\left(\frac{79}{15}-4 \gamma_E +2 \kappa \right)M \sigma +
\bigg(\frac{113-8 i   q-15 q^2}{15} +
2 \gamma_E \left(2\gamma_E-  \frac{79}{15}-2 \kappa \right)
-\frac{ \pi}{3}\left(\pi+\frac{19 i }{5} \right)
+\frac{79 \kappa }{15}\bigg) 2M^2\sigma ^2+
\nn \\   &
   \bigg(\frac{536213-252270 q^2}{23625}-\frac{296 i q}{45}-\frac{32 \gamma_E ^3}{3}
   -\frac{3002 i \pi }{225}+\frac{144 q
   \pi }{95}-\frac{158 \pi ^2}{45}+8\gamma_E  \left(\frac{i \left(8 q+19 \pi \right)}{15}-\frac{113}{15}+q^2+\frac{ \pi   ^2}{3}-\frac{79 \kappa }{15}\right)
    \nn \\   &
    +\left(\frac{103}{5}-q^2- \pi ^2-\frac{8 i q}{5}-\frac{19 i \pi }{5}\right) \frac{4\kappa}{3} +8\gamma_E ^2
   \left(\frac{79}{15}+2 \kappa \right)+\frac{8 \psi ^{(2)}(2)}{3}\bigg)M^3\sigma ^3 \Bigg] +o\left(\sigma^3\right),
      \end{align}
      \MC{$M$'s to be included above}
    \begin{align}	
    q^{\ph}&= i \pi  (1+\nu )+2   (1-\kappa -2 \log (4 \sigma ))M\sigma,
    \end{align}
    \MC{$M$'s to be included above}
    and
    \begin{align}\label{eq:q ell=1,m=1}
q(\sigma)&=-4 \pi M \sigma\Bigg[1+\left(\frac{109}{15}-4 \cdot\text{Eulerlog}\right) M\sigma +\bigg(\frac{64}{5}-\frac{8 i
   q}{15}-\frac{218 \cdot\text{Eulerlog}}{15}+4 \cdot\text{Eulerlog}^2
    -\frac{ \pi ^2}{3}\bigg)
2  M^2 \sigma ^2
\\   &
+ \bigg(\frac{1528463}{23625}-\frac{392 i q}{45}+\frac{872
   \cdot\text{Eulerlog}^2}{15}-\frac{32 \cdot\text{Eulerlog}^3}{3}-\frac{218 \pi ^2}{45}-\frac{8
   \kappa }{3} 
   -\frac{218 \kappa ^2}{15}+\frac{8 \kappa ^3}{3}+
   \nn \\   &
   \frac{2q^2}{3}
   \left(4 \kappa -\frac{3853}{175}\right)+
   \left(\frac{ \pi ^2}{3}-\frac{69}{5}+\frac{8 i q}{15}+q^2+\kappa   ^2\right)8\cdot\text{EulerLog}
   +\frac{8 \psi ^{(2)}(2)}{3}\bigg)M^3\sigma ^3\Bigg] +o\left(\sigma^3\right).\nn
\end{align}
\MC{$M$'s to be included above}
For the Wronskian, we find the expression 
\begin{align}\label{eq:check W fac ell=1,m=1}
	\check{W}^+\check{W}^-&=\frac{1}{144 \left(1-q^2\right) \left(2 q^2+2 i q \kappa -1\right)}\Bigg[1+
	  \left(4-2 i q-4 \gamma_E + \kappa -2 \psi\left(1+\frac{i q}{\kappa
	   }\right)\right)2M\sigma  \\
	   &+ \bigg(\frac{14503}{225}-36 i q-\frac{997 q^2}{75}+32 \gamma_E
	   ^2-\frac{76 i \pi }{15}-\frac{4 \pi ^2}{3}-8\gamma_E  \left(\frac{139}{15}-4 i q+2	   \kappa \right)	    +\left(8-\frac{17 i q}{3}\right) 2\kappa 
	\nn     \\	   &
+\left(4 i q-\frac{139}{15}+8	   \gamma_E -2 \kappa \ +2 \psi\left(1+\frac{i q}{\kappa }\right)\right)4\psi\left(1+\frac{i q}{\kappa }\right)
	   +\left(1+\frac{2 \kappa }{1-q^2}\right) 4\psi   ^{(1)}\left(1+\frac{i q}{\kappa }\right)\bigg)M^2\sigma ^2
	       \nn \\	   &
	   + \bigg(\frac{4}{225} 	   \left(18806-16126 iq-6837 q^2+2436 iq^3\right)
	   -\frac{256 \gamma_E ^3}{3}-\frac{40 i q
	   \log\left(4\sigma \kappa\right)}{57}-\frac{16 \pi}{285} (722 i+307 q) +\frac{8i\pi ^2 }{45} (41 i+30 q) \nn \\
	   &+\left(\frac{14203-11100 i q-4731 q^2}{75} -\frac{76 i   \pi }{5}-4 \pi ^2\right) \frac{2\kappa}{3} +64\gamma_E ^2 \left(\frac{79-30 i q}{15}+
	   \kappa \right) \nn \\
	   &+\frac{8\gamma_E }{3} \left(\frac{76 i \pi }{5}+4 \pi ^2-\frac{ 16783-9030 i q-2991  q^2}{75}-\frac{2}{5} (139-85 iq) \kappa
	   \right) \nn \\
	   &+\left(\frac{79}{15}-2 i q-4 \gamma_E + \kappa-\frac{2}{3} \psi\left(1+\frac{i   q}{\kappa }\right) \right) 16\psi\left(1+\frac{i q}{\kappa }\right)^2 \nn \\
	   &+\left(6-2 i q+\frac{\left(154-60 i q-15 q^2\right) \kappa }{15   \left(1-q^2\right)}-4\gamma_E  \left(1+\frac{2 \kappa }{1-q^2}\right)\right)8 \psi  ^{(1)}\left(1+\frac{i q}{\kappa }\right) \nn \\
	   &+4\psi\left(1+\frac{i q}{\kappa }\right)
	   \bigg(\frac{76 i \pi	   }{15}+\frac{4 \pi ^2}{3}-\frac{16783-9030 i q-2991 q^2}{225} -32 \gamma_E ^2 \nn \\
	   &-\frac{2}{15}  (139 -85 iq) \kappa +16\gamma_E  \left(\frac{79-30 i q}{15}	  + \kappa \right)-4\left(1+\frac{2 \kappa }{1-q^2}\right) \psi
	   ^{(1)}\left(1+\frac{i q}{\kappa }\right)\bigg)\nn \\
	   &+\frac{16 \psi
	   ^{(2)}(2)}{3}-\left(\frac{4-q^2}{3}+ \kappa \right) \frac{8  }{1-q^2}\psi ^{(2)}\left(1+\frac{i q}{\kappa }\right)\bigg)M^3\sigma ^3\Bigg]+o\left(\sigma^3\right), 
	   \nn
	   \end{align}
\MC{$M$'s to be included above}
and
	   \begin{align}\label{eq:ph W fac ell=1,m=1}
	  (W^+W^-&)^{\ph}=\frac{i q (1+\kappa +2 \log (\kappa ))}{1+\kappa }+2\nu  ( i \pi - \log (4M \kappa \sigma
   ))-4 \sigma  \log (4M \kappa \sigma ).
\end{align}
\MC{$M$'s to be included above}
We note that the  group of polygamma functions comes from the expansions of the $\Gamma$-functions which appear in $K_\nu$ and $K_{-\nu-1}$. 
We can now combine the above to give our important branch cut strength to Wronskian ratio
\begin{align}
	\frac{q(\sigma)}{W^+W^-}&=-\frac{M \pi  (q-i \kappa )^2 M\sigma }{36 \kappa ^2}(4 M\sigma\kappa)^2 e^{-\frac{i q}{1+\kappa}(1+\kappa+2\log\kappa)+2 M \sigma(1-\kappa+2\log\kappa)}\bigg[1+  \left(-\frac{41}{15}+4 \gamma_E +4 i q+4 \psi\left(1+\frac{i q}{\kappa }\right)\right)M \sigma \nn \\
	&+\Bigg(-\frac{7252}{225}+8 \gamma_E ^2+\frac{2 \pi ^2}{3}+\gamma_E  \left(-\frac{4}{5}+16 i q\right) -8 i q+\frac{10 i q \kappa
   }{3}+\frac{353 \kappa ^2}{75}+\frac{76}{15} \log (4M \kappa \sigma ) \nn \\
   &+\left(-\frac{88}{15}+16 \gamma_E +16 i q\right) \psi \left(1+\frac{i
   q}{\kappa }\right)+8 \psi\left(1+\frac{i q}{\kappa }\right)^2+\left(-4-\frac{8}{\kappa }\right) \psi ^{(1)}\left(1+\frac{i q}{\kappa
   }\right)\Bigg)M^2 \sigma ^2 \nn \\
   & +\frac{1}{15} \Bigg(-78 \pi ^2+\frac{8}{3} \gamma  \left(-881+15 \gamma  (7+4 \gamma )+15 \pi ^2\right)-128 q^2 \kappa +\frac{1412 \gamma 
   \kappa ^2}{5}+\frac{2006587-4413 \kappa ^2}{1575}\nn \\
   &+\frac{2}{15} i q \left(300 \pi ^2-258 q^2+300 \gamma  (6+12 \gamma +5 \kappa )-25 (580+41
   \kappa )\right)+24 (-1+20 \gamma +20 i q) \psi \left(1+\frac{i q}{\kappa }\right)^2\nn \\
   &+160 \psi\left(1+\frac{i q}{\kappa }\right)^3+\log(4\sigma
   \kappa) \left(-\frac{3116}{15}+304 \gamma +360 i q+304 \psi \left(1+\frac{i q}{\kappa }\right)\right)\nn \\
   &-\frac{4 (-44-41 \kappa +60
   \gamma  (2+\kappa )+60 i q (2+\kappa )) \psi ^{(1)}\left(1+\frac{i q}{\kappa }\right)}{\kappa }+\psi\left(1+\frac{i q}{\kappa }\right)
   \Bigg(\frac{4}{5} (-2677+600 \gamma ^2+50 \pi ^2 \nn \\
   &+80 \gamma  (4+15 i q)+353 \kappa ^2+50 i q (-3+5 \kappa ))-\frac{240 (2+\kappa )
   \psi ^{(1)}\left(1+\frac{i q}{\kappa }\right)}{\kappa }\Bigg)-40 \psi ^{(2)}(2) \nn \\
   &+\frac{40 (3+\kappa  (3+\kappa )) \psi ^{(2)}\left(1+\frac{i
   q}{\kappa }\right)}{\kappa ^2}\Bigg)M^3 \sigma ^3 \bigg] + o(\sigma^7)
\end{align}

\subsection{Mode $\ell=1$, $m=0$}

We obtain the following high order expansion for the transmission coefficient for the mode  $\ell=1$, $m=0$:
\begin{align}\label{eq:Btra ell=1,m=0}
  \check{B}^{\text{tra}}&=1+M \kappa  \sigma +\frac{3416 M^2 \kappa ^2 \sigma ^2}{9025}+\frac{M^3 \kappa  \left(-21660+1991 \kappa ^2\right) \sigma ^3}{9025}+o(\sigma^3),	\\
	   \Btrans_{\ph}&=  (1+\kappa +2 \log (\kappa ))M \sigma.
\end{align}
The expansion for the branch cut strength that we obtain is
\begin{align}\label{eq:check q ell=1,m=1}
	\check{q}(\sigma)&=-4\pi M\sigma\Bigg[1+\left(\frac{79}{15}-4 \gamma_E +2 \kappa \right)M \sigma +2\left(\frac{113-19 i \pi+79 \kappa}{15}-q^2+4 \gamma_E ^2-\frac{\pi ^2}{3}-2\gamma_E 
   \left(\frac{79}{15}+2 \kappa \right)\right)M^2 \sigma ^2 \nn \\
   &+ \Bigg(\frac{536213-252810 q^2}{23625}-\frac{32
   \gamma_E ^3}{3}+8\gamma_E  \left(\frac{19 i \pi -113-79 \kappa}{15}+q^2+\frac{ \pi ^2}{3}\right)
 \nn \\
   &+2\left( 4  \gamma_E ^2-\frac{19i\pi}{15}  - \frac{  \pi ^2}{3} \right)  \left(\frac{79}{15}+2  \kappa\right)+\frac{4\kappa}{3}\left(\frac{103}{5}-   q^2\right)  
 +\frac{8 \psi ^{(2)}(2)}{3}\Bigg)M^3\sigma ^3\bigg]+o(\sigma^3),
 \end{align}
 \MC{$M$'s to be included above}
together with
  \begin{equation}\label{eq:ph q ell=1,m=1}
   q^{\ph}= i \pi  (1+\nu )+2   (1-\kappa -2 \log (4 M\sigma ))M\sigma.
\end{equation}
\MC{$M$'s to be included above}
The expansion for the Wronskian factor in the expression for the Green function discontinuity along the cut is
\begin{align}\label{eq:check W fac ell=1,m=0}
\check{W}^+\check{W}^-&=144 \bigg[1+
\bigg(4-2 \gamma_E +\frac{ \left(1+q^2\right) \kappa }{1-q^2}\bigg) 2M\sigma 
+\Bigg(\frac{14503-12226 q^2+423 a^4}{225(1- q^2)}+8 \gamma_E
   ^2-4\frac{19 i \pi    + 139  \gamma_E}{15}+
       \nn\\   &
   \frac{(2-\gamma_E) 8 \left(1+q^2\right) \kappa }{1-q^2}
    +\frac{2\pi ^2}{3} \left(\frac{2 \kappa }{1- q^2}-1\right)
   \Bigg)M^2 \sigma ^2
   + \Bigg(\frac{8 \left(468   q^4-7171 q^2+9403\right)}{225 \left(1-q^2\right)}-\frac{32 \gamma_E ^3}{3}\nn\\
   &+\frac{2 \left(123 q^6-12163 q^4+1437 q^2+14203\right) \kappa }{225   \left(1-q^2\right)^2}
   +\frac{4\pi ^2 \left(8+52 q^2+\left(124-15 q^2\right) \kappa\right)}{45(1-q^2)}
   +16\gamma_E ^2   \left(\frac{79}{15}+\frac{ \left(1+q^2\right) \kappa }{1-q^2}\right)
   \nn\\   &
   -\frac{152 i\pi}{15}  \left(4 +\frac{ \left(1+q^2\right) \kappa   }{ 1-q^2}\right)
   +\frac{\gamma_E}{3}  \Bigg(\frac{304 i \pi   }{5}-\frac{4 \left(16783-14506 q^2+423 q^4\right)}{75 \left(1-q^2\right)}-\frac{1112 \left(1+q^2\right) \kappa }{5 \left(1-q^2\right)}
   \nn \\
   &+8\pi ^2 \left(1-\frac{2 \kappa }{  1-q^2}\right)\Bigg)-\left(\frac{4-q^2}{3}+ \kappa\right)\frac{ 8\psi   ^{(2)}(1)} {1-q^2}
   +\frac{16 \psi ^{(2)}(2)}{3}\Bigg)M^3\sigma ^3\Bigg]+o(\sigma^3),
    \end{align}
    \MC{$M$'s to be included above}
   together with
  \begin{equation}\label{eq:ph W fac ell=1,m=0}
   (\check{W}^+\check{W}^-)_{\text{ph}}=2\nu  ( i \pi - \log (4M \kappa \sigma   ))-4M \sigma  \log (4M \kappa \sigma ).
\end{equation}
\MC{$M$'s to be included above}

\subsection{Modes $\ell\geq2$}

\MC{Are these expansions valid both for $m\neq 0$ and $m=0$? Probably not}

One can perform the same analysis as in the previous two subsections in order
to arrive at expressions which are valid for general values of $\ell$ and $m$. The only caveat is 
that in order to obtain higher order terms in the expansion we must increase the minimum $\ell$ value for which the `general' $\ell$
expansion is valid. 
In the case of five leading orders, as here, the  `general' $\ell$ and $m$ expansion is valid for $\ell\geq 2$ \MC{Check}.
For the transmission coefficient we obtain the expression:
\begin{align}\label{eq:check Btra ell>1}
    \check{B}^{\text{tra}}&=1+ \kappa  M\sigma + \left(i m q \kappa +\frac{\left(4 \ell^4+8 \ell^3-3 \ell^2-7 \ell+4\right) \kappa ^2-\left(4 \ell^2+4 \ell-7\right) m^2 q^2}{(2 \ell+3)(2 \ell-1)}\right)
     \frac{M^2\sigma ^2}{(2 \ell+3)(2 \ell-1) } \nn \\
   &-\Bigg(\frac{6 \left(3 \ell^2+3 \ell-2\right) \kappa }{(\ell+1)\ell }+\frac{\left(4 \ell^2+4 \ell-7\right) m^2 q^2 \kappa }{(2 \ell+3)(2 \ell-1)}+
   \frac{\left(4 \ell^4+8 \ell^3+\ell^2-3 \ell-3\right) \kappa ^3}{3(2 \ell+3)(2 \ell-1)}\nn \\
   &+i m q \left(\frac{4  \left(4 \ell^6+12   \ell^5-11 \ell^4-42 \ell^3+\ell^2+24 \ell-9\right)}{(2 \ell+3)(2 \ell-1) (\ell+1)^2\ell^2}- 
   \kappa ^2\right)\Bigg)\frac{M^3  \sigma ^3}{(2 \ell+3)(2 \ell-1)}+o(M^3 \sigma^3),
   \end{align}
  and 
  \begin{equation}\label{eq:ph Btra ell>1}
  	   \Btrans_{\ph}=  (1+\kappa +2 \log (\kappa ))M \sigma-\frac{i m q (1+\kappa +2 \log (\kappa ))}{2 (1+\kappa )}.
\end{equation}
For the branch cut  strength we obtain
\begin{align}\label{eq:check q ell>1}
&	\check{q}(\sigma)=
	-4\pi M\sigma\Bigg[1+  \left(\frac{P_N}{P_D}+2 \kappa +4 \psi(1+\ell)\right)M\sigma+ \Bigg(1- q^2 \nn
	-\frac{2 i q m \left(8 \ell^2+7 \ell-7\right) }{(1+\ell)   P_D}+\frac{P_N}{P_D}\left( \kappa -i\pi\right)-\frac{ \pi ^2}{3} \nn \\ 
   &+\left(\frac{ P_N}{P_D}+2 \kappa   \right) 2\psi(1+\ell)+4 \psi(1+\ell)^2\Bigg)2 M^2\sigma ^2 
   + \Bigg(\frac{ 15 \ell^4+66 \ell^3 +58 \ell^2-17   \ell-12}{\ell (1+\ell) P_D}-\pi ^2 \left(\frac{  P_N}{P_D}+\frac{2 \kappa }{3}\right)
    \nn \\   &+q^2   \left(\frac{6 m^2}{ (2 \ell+3)(2 \ell-1) (\ell+1) \ell   }-\frac{ 15 \ell^2+19 \ell-9}{P_D}-\frac{2\kappa }{3}\right)+\frac{2 \kappa }{3}
     -\frac{2iq m}{(\ell+1) P_D} \left(\frac{  P_N}{  \ell }+2 \kappa\left(8   \ell^2+7 \ell-7\right)  \right)+ \nn \\
   &\frac{\pi}{P_D}  \left(\frac{4 am \left(5   \ell^2+5 \ell-3\right) }{\ell (\ell+1)} 
   -i P_N \left(\frac{  P_N}{P_D}+2  \kappa\right)\right)+\left(1-q^2-\frac{2 i qm \left(8   \ell^2+7 \ell-7\right) }{(\ell+1) P_D} 
-\frac{ \pi ^2}{3}  
+\frac{P_N    }{P_D}  \left(\kappa- i  \pi  \right)\right)4 \psi(1+\ell) \nn \\
   &+\left(\frac{P_N}{P_D}+2 \kappa \right) 4\psi(1+\ell)^2+\frac{4\left(4 \psi(1+\ell)^3+ \psi ^{(2)}(1+\ell)\right)}{3}
   +\frac{4 P_N \psi ^{(1)}(1+\ell)}{P_D}\Bigg)2M^3\sigma ^3\Bigg]+o\left(\sigma^4\right), 
      \end{align}
   and
   \begin{align}\label{eq:ph q ell>1}
 q^{\ph}&=i \pi  (1+\nu )+2   (1-\kappa -2 \log (4 M\sigma ))M\sigma,
\end{align}
      where 
      \begin{equation}
     P_N\equiv 15 \ell^2+15 \ell-11, \quad P_D\equiv  (2 \ell+3)  (2 \ell+1)(2 \ell-1).
      \end{equation}
For the Wronskian, we obtain 
\begin{align}\label{eq:W fac ell>1}
W^+W^-=&(\check{K}_\nu)^2e^{i q m \left(\frac{2 \log (\kappa )}{\kappa +1}+1\right)}\left(4 \kappa M\sigma\right)^{-2\ell}\bigg(1-\left(2\log (4M \kappa \sigma)-4\psi(\ell+1)\right)
2M\sigma+\bigg(2 \log^2 (4M \kappa \sigma ) \nonumber \\
&\left.\left.-\left(\frac{15 \ell^2+15 \ell-11}{(2 \ell+3) \left(4 \ell^2-1\right)}+8 \psi(\ell+1)\right) \log (4M \kappa \sigma )+8
   \psi(\ell+1)^2\right) 4M^2\sigma^2+o(\sigma^2)\right).
\end{align}
\MC{This is the only quantity not given up to $O(\sigma^3)$, can we not obtain that order?}
\MC{Do the above W's need overhead bars?}
Where $K_\nu$ is
\MC{Note I've tidied up the expression below. So if you write in a higher order, please don't mess the lower orders currently written}
\begin{align}
	\check{K}_\nu&=\frac{\Gamma(2 \ell+1) \Gamma (2+2 \ell) \Gamma \left(1+\frac{i q m}{\kappa }\right)}{\Gamma^2(\ell+1) \Gamma \left(1+\ell+\frac{i q m}{\kappa }\right)}\Bigg[1-  \left(\kappa +\frac{2 (1+\kappa ) \psi\left(1+\frac{i q m}{\kappa }\right)}{\kappa }-\frac{2 \psi\left(1+\ell+\frac{i q
   m}{\kappa }\right)}{\kappa }\right)M\sigma \nn \\
   &+ \Bigg(\frac{\left(8 \ell^4+16 \ell^3-4 \ell^2-13 \ell+4\right) \kappa ^2+2   (2 \ell+1) m^2 (\kappa^2-1)}{(2 \ell-1)^2 (2 \ell+3)^2}
   -
   \frac{i q m \left((16+\kappa)(\ell+1)\ell  -12 \right)}{\ell    (2 \ell+3)(2 \ell-1)(\ell+1) } \nn \\
   &+2\frac{2 \left(15 \ell^2+15 \ell-11\right) \left( \psi(2 \ell+2)+ \psi(2 \ell+1)-\psi(\ell+1)\right)
   -\left(8 \ell^3+27 \ell^2+13 \ell-14\right) \psi\left(1+\ell+\frac{i a   m}{\kappa }\right)
   }{ (2 \ell+3)(2 \ell+1)(2 \ell-1) }
    \nn \\
   &
   +\frac{2}{\kappa ^2}\left( (1+\kappa   ) \psi\left(1+\frac{i q m}{\kappa }\right)- \psi\left(1+\ell+\frac{i q m}{\kappa }\right)\right)^2
   +2 (1+\kappa )\psi\left(1+\frac{i q m}{\kappa }\right)
       \nn \\   &
   -4 \psi ^{(1)}(1+\ell)+
   \frac{2}{\kappa ^2}\left((1+\kappa )^2 \psi ^{(1)}\left(1+\frac{i q m}{\kappa }\right)- \psi   ^{(1)}\left(1+\ell+\frac{i q m}{\kappa }\right)\right)
   \Bigg)M^2\sigma ^2\Bigg]+o(\sigma^2),
\end{align}
with the phase as in Eq.~(\ref{Eq:KnuPh}) and the term proportional to $K_{-\nu-1}$ will only appear at the next order.

We also obtain the leading order for $K_{-\nu-1}$.
As a first step, we obtain
\begin{align}\label{eq:K-nu-1 lead}
K_{-\nu-1}\sim
-\frac{2^{\ell-1}\kappa^{s+\ell+1}\Gamma^2(\ell+s+1)}{\nu_2^2\Gamma(2\ell+1)\Gamma(2\ell+2)}
\frac{\Gamma(1-s-2i\epsilon_+)}{\Gamma(-\ell-i\tau-\nu_2\epsilon^2)}
\epsilon^{s+\ell-1},\quad \epsilon\to 0,
\end{align}
where $\nu_2$ is given in Eq.(\ref{eq:nu2}).
The last $\Gamma$-fraction in Eq.(\ref{eq:K-nu-1 lead}), $\Gamma(1-s-2i\epsilon_+)/\Gamma(-\ell-i\tau-\nu_2\epsilon^2)$ requires some care.
By inspection, we observe that it is $O(1)$ if  $m\neq 0$  and/or $s\ge 1$.
Otherwise, however, it goes like
\begin{equation}
\frac{\Gamma(1-s-2i\epsilon_+)}{\Gamma(-\ell-i\tau-\nu_2\epsilon^2)}
\sim \frac{(-1)^{\ell+1}i\Gamma(1-s)\Gamma(\ell+1)}{\kappa}\epsilon, \quad \epsilon\to 0,\quad \text{if}\ m=0\ \text{and}\ s<1.
\end{equation}
This means that the leading order asymptotics
of  $K_{-\nu-1}$ when   $m\neq 0$  and/or $s\ge 1$ are already manifest in Eq.(\ref{eq:K-nu-1 lead}), with the last $\Gamma$-fraction of $O(1)$ .
Otherwise, the asymptotics are of one order higher in $\epsilon$ and they are manifest in
\begin{align}\label{eq:K-nu-1 lead m=1,s<1}
K_{-\nu-1}\sim
\frac{(-1)^{\ell}2^{\ell-1}\kappa^{s+\ell}\Gamma^2(\ell+s+1)\Gamma(1-s)\Gamma(\ell+1)}{\nu_2^2\Gamma(2\ell+1)\Gamma(2\ell+2)}
\epsilon^{s+\ell},\quad \epsilon\to 0,\quad \text{if}\ m=0\ \text{and}\ s<1.
\end{align}
For completeness, we have obtained the leading order result for $K_{-\nu-1}$ for general spin $s$.
Let us focus the following discussion on the case $s=0$ and $\ell\ge 2$.
We have  seen  above that, $e^{K_{\nu}^{\ph}}$  captures the leading-order of $K_{\nu}$, so that
  $K_{\nu}\sim \epsilon^{-\ell}$ for small $\epsilon$.
On the other  hand, $K^{\ph}_{-\nu-1}$  captures the leading order of $K_{-\nu-1}$, so that  $K_{-\nu-1}\sim \epsilon^{\ell}$,
only when $m=0$; when $m\neq 0$, we have at  $K_{-\nu-1}\sim \epsilon^{\ell-1}$.
 

\section{Low-frequency Expansion of the Radial Solutions}\label{sec:low freq r-dep}

\begin{figure}[htb!]
\centering
\includegraphics[width=10cm]{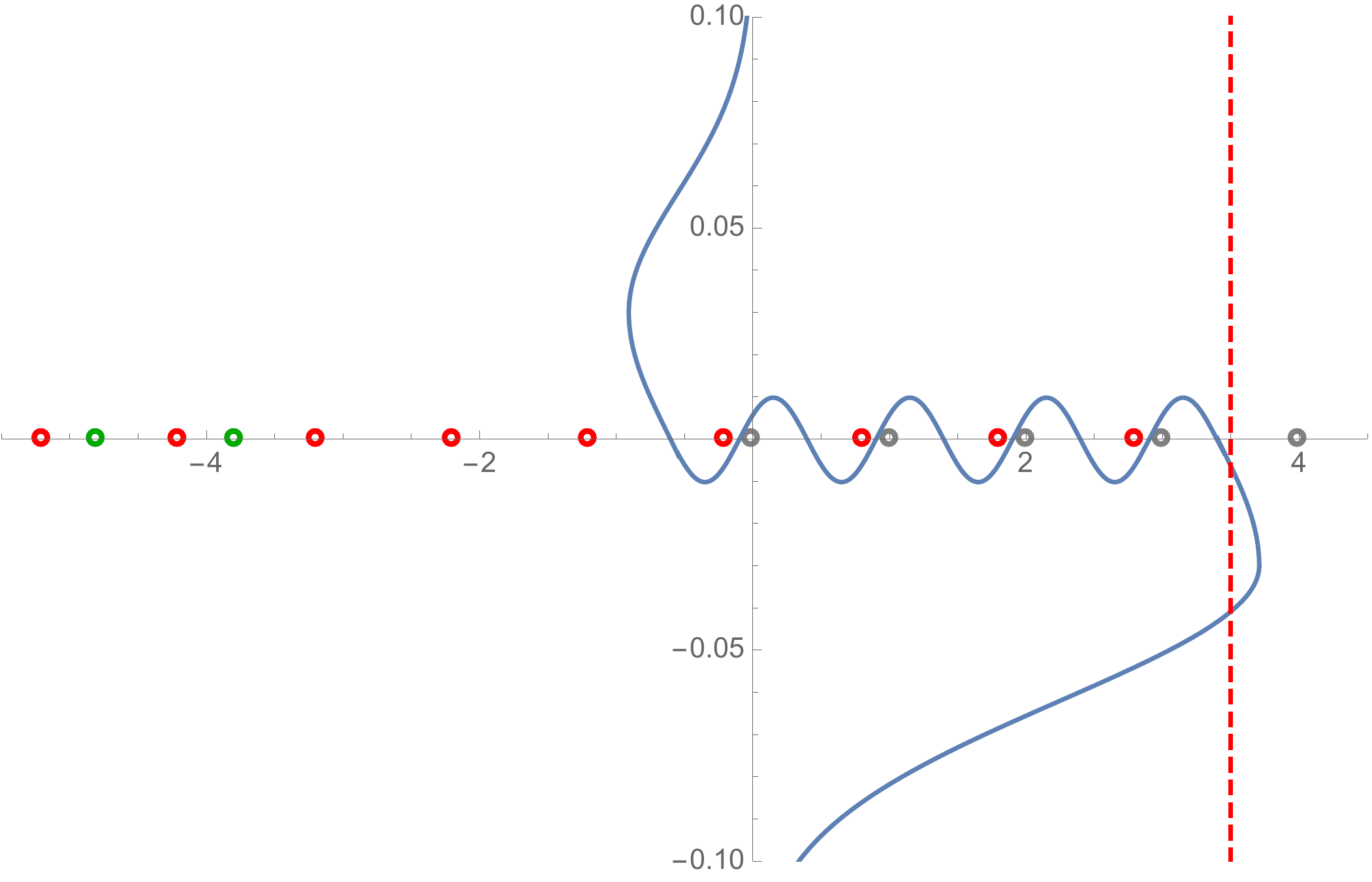}
\caption{
Plot of the contour deformation in the
complex $t$-plane for the  hypergeometric function Eq.(\ref{eq:Mellin-Barnes}).
The straight blue curve is the contour (C1). The dashed red vertical line is the
contour (C2). The circled crosses correspond to the poles of the various $\Gamma$ functions in the integrand:
the blue ones are those of $\Gamma(-t)$; the green ones are those of $\Gamma(a+t)$; the red
ones  are those of $\Gamma(b+t)$.
}
\label{fig:Barnescontour}
\end{figure}

In the expression in Eq.(\ref{Eq:GBC}) for the branch cut modes, the only radial solution that is required is the ingoing radial solution.
In this section, we provide a low-frequency expansion for the ingoing radial solution which is valid at arbitrary radius, so that it is valid in the strong field as well.
Our starting point is the representation in Eq.(\ref{Eq:Rin 2F1}).
One of the ingredients in this representation are the series coefficients $\an{n}$, for which we give a low-frequency expansion in App.\ref{sec:MSTexpansions}.
In order to complete the low-frequency expansion of the ingoing solution, we also need to expand out the hypergeometric functions themselves.
 A representation of the  hypergeometric functions  that is useful for expanding for low frequency at arbitrary radius
  is  the Mellin-Barnes integral found in \cite{bk:AS}. This gives the functions as an integral
\begin{equation}\label{eq:Mellin-Barnes}
\vphantom{}_2F_1(a,b;c;x)=\frac{\Gamma(c)}{\Gamma(a)\Gamma(b)}\frac{1}{2\pi i}\int_{\text{C1}}
\frac{\Gamma(a+t)\Gamma(b+t)\Gamma(-t)}{\Gamma(c+t)}(-x)^t\text{d}t,
\end{equation}
where the integral goes along the (red) contour C1 in Fig \ref{fig:Barnescontour}, separating the poles of $\Gamma(a+t)$ and $\Gamma(b+t)$ from those of $\Gamma(-t)$. In the figure,
 the poles of  $\Gamma(-t)$ (seen in green) are at $t\in  \mathbb{N}$ \MC{But on the figure they're above the real line?}.
 The functions  $\Gamma(a+t)$ and $\Gamma(b+t)$ have poles (shown in red and black) occuring at, respectively,
\begin{eqnarray}
t&=&-a,-a-1,-a-2... \\
t&=&-b,-b-1,-b-2...
\end{eqnarray}
For the case of interest to us, Eq.(\ref{Eq:Rin 2F1}), we have the following values:
\begin{equation}
a=n+\nu+1-i\tau, \quad b=-n-\nu-i\tau,\quad c=1-s-i\epsilon-i\tau.
\end{equation}
Practically speaking, evaluating the integral in Eq.(\ref{eq:Mellin-Barnes}) is at first a tricky prospect: to even calculate it numerically, specifying the contour would require a lot of care. It can be made easier by shifting to the contour C2 (in blue), while picking up some of the residues of the poles of $\Gamma(-t)$, shown in green. 
Changing contour splits the calculation into two workable pieces. First there is the contribution from the residues: 
\begin{equation}\label{eq:Fres}
F^{\rm res}_n=\sum_{k=0}^{N}\frac{(-1)^k}{k!}\frac{\Gamma(a+k)\Gamma(b+k)}{\Gamma(c+k)}(-x)^k
\end{equation}
where $n$ indicates the dependence on the summation index in Eq.(\ref{Eq:Rin 2F1}) and
 $N$ is the number of poles of $\Gamma(-t)$ that we pass through in changing contours from C1 to C2. This expression  allows a Taylor expansion in $\epsilon$ to obtain exact expressions for the contributions at each order. What is left is the integral, now over a much simpler contour   (C2) of our choosing that approches complex infinity on either side, passing somewhere in between the $N$th and $(N+1)$th pole. A change of variables $t=i y +N+\frac{1}{2}$ gives the integral as
\begin{equation}
F^{\rm int}_n=\frac{1}{2\pi}\int_{- \infty}^{\infty}
\frac{\Gamma\left(a+i y +N+\frac{1}{2}\right)\Gamma\left(b+i y +N+\frac{1}{2}\right)}{\Gamma\left(c+i y +N+\frac{1}{2}\right)}\Gamma\left(-i y -N-\frac{1}{2}\right)(-x)^{i y +N+\frac{1}{2}}\text{d}y. \label{Eq:BarnesInt}
\end{equation}
In summary, we have obtained the expression
\begin{equation}\label{eq:Rin=pin} 
\Rin=e^{i \epsilon \kappa x}(-x)^{-s-i(\epsilon+\tau)/2}(1-x)^{i(\epsilon-\tau)/2}
p_{\text{in}}^\nu(x),
\end{equation}
with 
\begin{equation}
p_{\text{in}}^\nu=
\sum_{n=-\infty}^{\infty}\an{n} \left(F^{\rm res}_n+F^{\rm int}_n\right),
\end{equation}
which is readily amenable to a low frequency expansion.
A closed-form expression for the integral contribution here is, however, not apparent. We are forced to expand the integrand and leave each order as a  integral which we can evaluate
numerically to any desired accuracy. 
We then need to sum the residues and evaluate the integrals for each $n$. We truncate the $n$-sum when the leading $\epsilon$ contribution of the terms neglected is smaller 
than the order in $\epsilon$ required, i.e. for an expansion to $\epsilon^3$ we can ignore terms with $|n|>3$\MC{Check this is true for$\ell=0$}. 
In the next subsections we give explicit expansions 
for the lowest modes, $\ell=0$ and $ 1$.


\subsection{Mode $\ell=m=0$}

Each of the hypergeometric functions in Eq.(\ref{Eq:Rin 2F1}) must be computed using the Barnes integral method described above, where the number 
of terms in the infinite sum in $n$ is bound by the low frequency expansion. Working to $o(\sigma^3)$, we find, for $\ell=m=0$,
\begin{align}\label{eq:pin ell=0}
p_{\text{in}}^\nu=&	\frac{7}{9}\bigg(1+\bigg[\left(\frac{1}{3}-2 x\right) \kappa+\frac{2}{\kappa}\int_{-\infty}^\infty \frac{ (-x)^{\frac{1}{2}+i y}  \text{sech}(\pi  y)}{(1+2 i y)  }dy \bigg]M \sigma+\bigg[\frac{2056}{2205}-\frac{592 \kappa ^2}{2205}+ \left(-2-4 \kappa -\frac{4 \kappa ^2}{3}\right)x \nonumber\\
&\quad +\frac{8 x^2 \kappa ^2}{3}+\frac{2}{\kappa}\int_{-\infty}^\infty \frac{ (-x)^{\frac{1}{2}+i y}  \text{sech}(\pi  y)}{(1+2 i y)  }  \left(-\frac{2 H_{i y-\frac{1}{2}}}{\kappa }+2 H_{i y+\frac{1}{2}}+3 \kappa -\frac{\kappa  x (2 y-5 i)}{2 y-3 i}\right)\bigg] M^2 \sigma^2\bigg)+o(M^2\sigma^2),
\end{align}
\MC{Above there're $M$'s but in the expressions in the previous section we had $M=1$. Change above $\sigma\to M\sigma$ and $a\to q$}
where $H_c$ is the complex harmonic number.


\subsection{Mode $\ell=1$}

The expressions for the $\ell=1$ case are, algebraically, significantly more complicated than those for $\ell=0$, particularly when $m=1$. The main source of this is that the arguments of the hypergeometric function are no longer integers when $\epsilon=0$. We write
\begin{align}
p_{\text{in}}^\nu&= p_{\text{in}}^\text{res}+p_{\text{in}}^\text{int},
\end{align}
where we have defined
\begin{align}
	p_{\text{in}}^\text{res}&\equiv\sum_{n=-\infty}^{\infty}a_n^{\nu} F^{\rm res}_n, \qquad	p_{\text{in}}^\text{int}\equiv\sum_{n=-\infty}^{\infty}a_n^{\nu} F^{\rm int}_n.
\end{align}
Then, to order $O(M^2 \sigma ^2)$ with $m=0$,
\begin{align}\label{eq:pinres ell=1,m=0}
p_{\text{in}}^\text{res}&=(1-2x)-M \left(2x \left(2+\frac{3}{\kappa }\right)-(1-2x)^2 \kappa \right) \sigma +M^2 \bigg(-\frac{199}{30}+\frac{3 (1-2x)^3 \kappa
   ^2}{5}+(1-2x)^2 \left(\frac{23}{6}+3 \kappa \right) \nn \\
   &+(1-2x) \left(\frac{14}{5}+\frac{4}{\kappa ^2}+\frac{10}{\kappa }-2 \kappa -\frac{1999 \kappa
   ^2}{9025}\right)-\frac{4+10 \kappa +\kappa ^3}{\kappa ^2}\bigg) \sigma ^2+o(\sigma^2),
   \end{align}
   and
   \begin{align}\label{eq:pinint ell=1,m=0}
p_{\text{in}}^\text{int}&=-\int_{-\infty}^{\infty}\frac{2 i M (-x)^{\frac{3}{2}+i y} (-5 i+2 y) \text{sech}(\pi  y)}{(-i+2 y) (-3 i+2 y) \kappa } \Bigg(\sigma+2 M \bigg[\frac{2}{\kappa }+\frac{\gamma_E  (-1+\kappa )}{\kappa }+\frac{19 \kappa }{30}+\frac{(3 i-2 y) \kappa }{6 x (-5 i+2 y)} \nn \\
&+\frac{x (63-4 y (-8 i+y))
   \kappa }{3 (5 i-2 y)^2}-\frac{\psi\left(\frac{1}{2}+i y\right)}{\kappa }+\left(1+\frac{1}{\kappa }\right) \psi\left(\frac{5}{2}+i y\right)-\frac{\psi\left(\frac{7}{2}+i y\right)}{\kappa }\bigg]\sigma^2\Bigg)dy+o(\sigma^2).
\end{align}

For $m=1$:
\begin{align}
	p_{\text{in}}^\text{res}&=1-2 \left(1-q^2\right) x+\frac{i q \left(3-2 q^2\right) x \kappa }{1-q^2}+M \Bigg(\frac{25 i q \left(-1+q^2\right)}{361
   \left(1-q^2\right)}-\frac{i q \left(46-81 q^2+36 q^4\right) x^2}{6 \left(1-q^2\right)}+\kappa +\frac{\left(8-21 q^2+12 q^4\right) x^2 \kappa
   }{2-2 q^2}\nn \\
   &+x \left(2 i q\bigg(\frac{ \left(5851-11341 q^2+5490 q^4\right)}{1083 \left(1-q^2\right)}+ \left(5-4 q^2\right) \kappa \bigg)-2 \left(2-7 q^2+4 q^4\right)-\frac{2 \left(5415-9679 q^2+5490 q^4\right) \kappa }{1083 \left(1-q^2\right)}\right)\Bigg) \sigma \nn \\
   & +M^2
   \Bigg(\frac{8 \left(462441-607268 q^2+144827 q^4\right)}{9774075 \left(1-q^2\right)}-\frac{4 \left(90-379 q^2+468 q^4-180 q^6\right) x^3}{75
   \left(1-q^2\right)}+\frac{958 i q \left(-1+q^2\right)^2 \kappa }{5415 \left(1-q^2\right)^2}\nn \\
   &-\frac{i q \left(-1596+4619 q^2-4464 q^4+1440
   q^6\right) x^3 \kappa }{150 \left(1-q^2\right)^2}+x^2 \Bigg(\frac{i q \left(-132+337 q^2-276 q^4+72 q^6\right)}{4-3 q^2} \nn \\
   &+\frac{i q
   \left(-839+2369 q^2-2274 q^4+744 q^6\right) \kappa }{25 \left(1-q^2\right)^2}+\frac{3380-10535 q^2+11772 q^4-4464 q^6}{150
   \left(1-q^2\right)}\nn \\
   &+\frac{2 \left(72-412 q^2+699 q^4-468 q^6+108 q^8\right) \kappa }{3 \left(4-7 q^2+3 q^4\right)}\Bigg) \nn \\
   &+x
   \Bigg(-\frac{313653406-1483629109 q^2+2354055315 q^4-1496850012 q^6+312770400 q^8}{9774075 \left(1-q^2\right)} \nn \\
   &-\frac{4 \left(30324-167171
   q^2+301001 q^4-223086 q^6+58932 q^8\right) \kappa }{1083 \left(4-7 q^2+3 q^4\right)}-\frac{4 i q \left(-75388+209933
   q^2-193620 q^4+58932 q^6\right)}{1083 \left(4-3 q^2\right)}\nn \\
   &+\frac{i q \left(662855204-2362316633 q^2+3039926241 q^4-1653235212 q^6+312770400
   q^8\right) \kappa }{9774075 \left(1-q^2\right)^2}\Bigg)\Bigg) \sigma ^2+o(\sigma^2).
\end{align}
\begin{align}
	p_{\text{in}}^\text{int}&=\int_{-\infty}^{\infty} \frac{(-x)^{-\frac{1}{2}-i y} \Gamma \left(\frac{i q}{\kappa }\right)}{\Gamma \left(\frac{1}{2}-i y\right) \Gamma \left(\frac{i q}{\kappa }+i
   y+\frac{1}{2}\right)}\Bigg[\frac{x \left(8 i q-6 i q^3-5 \kappa +6 q^2 \kappa +y \left(4 q \left(-1+q^2\right)+2 i \left(-1+2 q^2\right) \kappa \right)\right)}{\pi  (-3+4 y
   (-2 i+y)) \kappa } \nn \\
   &+\Bigg(-\frac{q+i \kappa }{-6 \pi  y+3 i \pi }+\frac{2 \left(q^3 (6 i-4 y)+4 q (-2 i+y)+(5+2 i y) \kappa +q^2 (-6-4 i y) \kappa \right)}{\pi  (-3+4 y (-2 i+y)) \kappa ^2}\Bigg(\frac{25 i q \kappa }{722}-\psi\left(-1+\frac{i q}{\kappa }\right) \nn \\
   &+\frac{\left(4 q^3 y-\left(25+4 y^2\right) \kappa  (1+\kappa )-4 q y
   (1+\kappa )^2+q^2 \left(-4+\left(21+4 y^2\right) \kappa \right)\right) \psi\left(1+\frac{i q}{\kappa }\right)}{-25-4 y^2+q^2 \left(21+4
   y^2\right)-8 q y \kappa }-\psi\left(2+\frac{i q}{\kappa }\right) \nn \\
   &+\psi\left(\frac{1}{2}+i y+\frac{i q}{\kappa
   }\right)+\frac{\left(-4 q^3 y+\left(25+4 y^2\right) \kappa  (1+\kappa )+4 q y (1+\kappa )^2+q^2 \left(4-\left(21+4 y^2\right) \kappa
   \right)\right) \psi\left(\frac{5}{2}+i y+\frac{i q}{\kappa }\right)}{-25-4 y^2+q^2 \left(21+4 y^2\right)-8 q y \kappa } \nn \\
   &+\psi
  \left(\frac{7}{2}+i y+\frac{i q}{\kappa }\right)\Bigg)x+\bigg(-\frac{q \left(-251+457 q^2-210 q^4+20 y^2-44 q^2 y^2+24 q^4 y^2\right)}{3 \pi  (-i+2 y) (-3 i+2 y) (-5 i+2 y) \kappa ^2}\nn \\
   &+\frac{2
   \left(63 i-176 i q^2+105 i q^4-32 y+108 q^2 y-72 q^4 y-4 i y^2+16 i q^2 y^2-12 i q^4 y^2\right)}{3 \pi  (-i+2 y) (-3 i+2 y) (-5 i+2 y) \kappa
   }\nn \\
   &+\frac{48 q y \kappa ^2}{\pi  (-1-2 i y) (-3 i+2 y) (-5 i+2 y)}\bigg)x^2 \Bigg)M\sigma\Bigg]dy.
\end{align}
\MC{$M$'s to be included above}
\MC{Where's the Barnes for $\ell=1$ and $m=- 1$?}


%


\section{Late-time Tail of the Green Function} \label{sec:late-time}

We now have all the ingredients for obtaining the late-time behaviour of the Green function.
We calculate $G_{BC}$ using Eq.(\ref{eq:G_BC}). Apart from the spin-weighted spheroidal harmonics, this requires the branch cut modes $\Disc \Glm$, which we obtain using Eq.(\ref{Eq:GBC}).
In its turn, this requires knowledge of the branch cut strength $q$, the Wronskian factor $\Wbp\Wbm$ and the radial `in' solutions $\Rinhat$.
We have given explicit  high-order expansions for small frequency for all these quantities for spin $s=0$.
A high order expansion for the branch cut strength is given in Eq.(\ref{eq:q ell=0}) for $\ell=0$, in Eqs.(\ref{eq:q ell=1,m=1}), (\ref{eq:check q ell=1,m=1}) and
(\ref{eq:ph q ell=1,m=1}) for $\ell=1$, and in Eqs.(\ref{eq:check q ell>1}) and (\ref{eq:ph q ell>1}) for $\ell \geq 2$.
A high order expansion  for the  Wronskian factor is given in Eqs.(\ref{eq:check W fac ell=0}) and (\ref{eq:ph W fac ell=0}) for $\ell=0$,
in Eqs.(\ref{eq:check W fac ell=1,m=1}), (\ref{eq:ph W fac ell=1,m=1}), (\ref{eq:check W fac ell=1,m=0}) and (\ref{eq:ph W fac ell=1,m=0}) for $\ell=1$,
and in Eq.(\ref{eq:W fac ell>1}) for $\ell\geq 2$.
An expansion to arbitrary order for small frequency for the radial functions at arbitrary radius can be readily obtained from Eq.(\ref{eq:Rin=pin}),
by expanding the residues Eq.(\ref{eq:Fres}) and the integrand in Eq.(\ref{Eq:BarnesInt}).
Explicit expansions  are given via Eq.(\ref{eq:pin ell=0}) for $\ell=0$ and  via Eqs.(\ref{eq:pinres ell=1,m=0}) and (\ref{eq:pinint ell=1,m=0}) for $\ell=1$.
We note that these radial solutions still have to be normalized by dividing by $\Btrans$ as per Eq.(\ref{eq:hatted slns}).
We give expansions for $\Btrans$ in Eqs.(\ref{eq:check Btra ell=0}) and (\ref{eq:ph Binc/ref/tra ell=0})  for $\ell=0$,
in Eqs.(\ref{eq:check Btra ell=m=1}), (\ref{eq:ph Btra ell=m=1}) and (\ref{eq:Btra ell=1,m=0}) for $\ell=1$,
and in Eqs.(\ref{eq:check Btra ell>1}) and (\ref{eq:ph Btra ell>1}) for $\ell\ge 2$.
The above quantities require expansions for the series coefficients $\an{n}$ and for $\nu$, which we give in App.\ref{sec:MSTexpansions}.

In the following subsection, we use the above prescription for obtaining explicit high-order expansions for the Green function at late times for $\ell=0$ and
$\ell=1$ for space-time points correcponding to a particular orbit in Kerr.
Before that, though, let us give a quick derivation of the leading power law tail decay of the Green function, where here we ignore any
frequency-independent factors in the modes and  time-independent factors in the final result.
From the equations referred to in the previous paragraph, we readily have: $q\sim \fNIA$, 
$\Wbp\Wbm \sim \fNIA^{-2\ell}$ and $\Rinhat \sim O(1)$.
Therefore, from Eq.(\ref{Eq:GBC}), we have $\Disc \Glm \sim \fNIA^{2\ell+2}$.
Trivially, we also have ${}_s\Slm(\theta)=O(1)$ \MC{Please check}.
Then, Eq.(\ref{eq:Disc Gell}) yields $\Disc G_{\ell} \sim t^{-2\ell-3}$ as the leading-order behavior at late times.

Expansions for small $a\omega$ for the spin-weighted spheroidal expansion are given in, e.g.,~\cite{Kavanagh:2016idg}.
\MC{Please check. Also, these are small $a\omega$, not $M\omega$? so the expansions we give below for the GF, etc are not for small $M\omega$ only but also small $a\omega$?
or is it that we do $a\omega=(M \omega)(a/M)$, then fix $a/M$, and so they effective are small $M\omega$ expansions?}
Using these, together with the above expansions for the radial part, we can calculate the  retarded Green function in the time domain. To do this we must  first specify a sorce point and a field
 point. We choose the points to lie on a timelike circular geodesic on the equator of radius $r_0=\frac{9 \sqrt{11}}{5}M$, so that $\varphi=\Omega t$, 
 with $\Omega=M^{1/2}/\left(r_0^{3/2}+aM^{1/2}\right)$.
  \MC{check $\Omega$}
 Our calculattion thus shows the contribution from a point at a fixed time in the past on a particle far into its orbit.  
 We choose a value of $a=6M/10$ for the angular momentum of the black hole.


\subsection{Mode $\ell=0$}


\MC{I changed the sign of $q$ wrt what there used to be everywhere above. However. I haven't carried over this change of sign to any of the below.
Please check whether the signs below agrees with the numerics or we need to change it (as I'd expect)? Sign  changed!}
Giving the results to six digits of accuracy, the branch cut mode for $\ell=0$ is
\begin{align}\label{eq:Disc Gell ell=0}
\Disc G_{\ell=0}(t)
=
\frac{8}{\tilde{t}^3}+\frac{123.074}{\tilde{t}^4}-\frac{1408. \log \left(\tilde{t}\right)-7254.94}{\tilde{t}^5}-\frac{57605.8 \log
   \left(\tilde{t}\right)-185111.}{\tilde{t}^6}+o\left(\tilde{t}^{-6}\right)
\end{align}
\MC{I have changed the previous notation $\Disc G_{\ell,m}(t)$ to $\Disc G_{\ell}(t)$; please check if what you calculated in Eq.(\ref{eq:Disc Gell ell=0}) is indeed 
$\Disc G_{\ell}(t)$ defined in Eq.(\ref{eq:Disc Gell})}
\MC{Include same number of digits for all the coefficients. Is the $8.$ exact?}
with $\tilde{t}=t/M$. At leading order this is identical to that of a Schwarzschild black hole. Given the poor convergence displayed by the increasingly large coefficients of the powers of $1/\tilde{t}$, these extra terms will offer dramatic improvement at increasingly earlier times.

\subsection{Mode $\ell=1$}

We are evaluating the Green function on the equator ($\theta=\pi/2$).
For $s=0$ and $\ell=1$ the spheroidal harmonic is zero on the equator when $m=0$, and so we do not need this mode.
Furthermore, we use the symmetry mentioned above, $\Disc G_{1,1}+\Disc G_{1,-1}=2 \text{Re}(\Disc G_{1,1})$ on the equator.
After summing over $m$ and using the same circular orbit field point as before, we find the time domain branch cut mode for $\ell=1$ to be, within six digits of accuracy,
\begin{align}\label{eq:Disc Gell ell=1}
\Disc G_{\ell=1}(t)
=&\cos(\Omega t )\left(\frac{801.921}{\tilde{t}^5}+\frac{5240.65}{\tilde{t}^6}+\frac{451508.-121892. \log \tilde{t}}{\tilde{t}^7}+\frac{432880.-1.11645  \log \tilde{t}}{\tilde{t}^8}\right) \nonumber \\
&+\sin(\Omega t)\left(\frac{2.63137}{\tilde{t}^5}-\frac{2398.86}{\tilde{t}^6}-\frac{38679.2+399.969 \log \tilde{t}}{\tilde{t}^7}-\frac{2790969 - 887701.  \log \tilde{t}}{\tilde{t}^8}\right)
+o\left(\tilde{t}^{-8}\right).
\end{align}
\MC{I have changed the previous notation $\Disc G(t)$ to $\Disc G_{\ell}(t)$; please check if what you calculated in Eq.(\ref{eq:Disc Gell ell=1}) is indeed 
$\Disc G_{\ell}(t)$ defined in Eq.(\ref{eq:Disc Gell})}
Here we split the integral into its two oscillatory parts coming from $e^{i m \phi}=e^{i m \Omega t}$.


\subsection{Comparisons with real frequency integration}

As a test of our method, we compare the results of the previous subsections with some specific cases of the 
Green function which we have calculated using an integration of the frequency domain Green function along the original real frequency
contour. In this calculation we generate the ingoing and upgoing homogenous solutions using the MST method, numerically
evaluating for each (real) frequency point and continuing the series expansion until we get convergence to a specified
accuracy goal. 

In this calculation we use the following procedure:
\begin{enumerate}
	\item We find the renormalised angular momentum $\nu$ using either root-finding methods or the monodromy methods of \cite{castro2013black}. We did this over a range $-10<=\epsilon<=10$ with a grid spacing of $0.001$.
	\item We calculate the homogenous solutions on this grid using the MST expansions \eqref{Eq:Rin 2F1} and \eqref{eq:Rup series U}. Here the series coefficients  $\an{n}$ 
	are evaluated using the continued fractions until a given tolerance. We truncate the sums over $n$'s when a specified accuracy goal is met. 
	\item From these solutions we form the retarded Green function modes as per \eqref{eq:rGF}. 
	\item By interpolating this function, the time-domain Green function is approximated by a numerical evaluation of the integral \eqref{Eq:TDGF} over the restricted $\epsilon$ range.\MC{No word about spheroidals?} Here we introduce two error functions, which serve to smoothly cut off the integral at its two limits giving a better approximation to the full integral (a justification for a similar smoothing in the $\ell$-sum is given in~\cite{CDOW13}). 
\end{enumerate}

\begin{figure}[htb!]
\includegraphics[scale=.4]{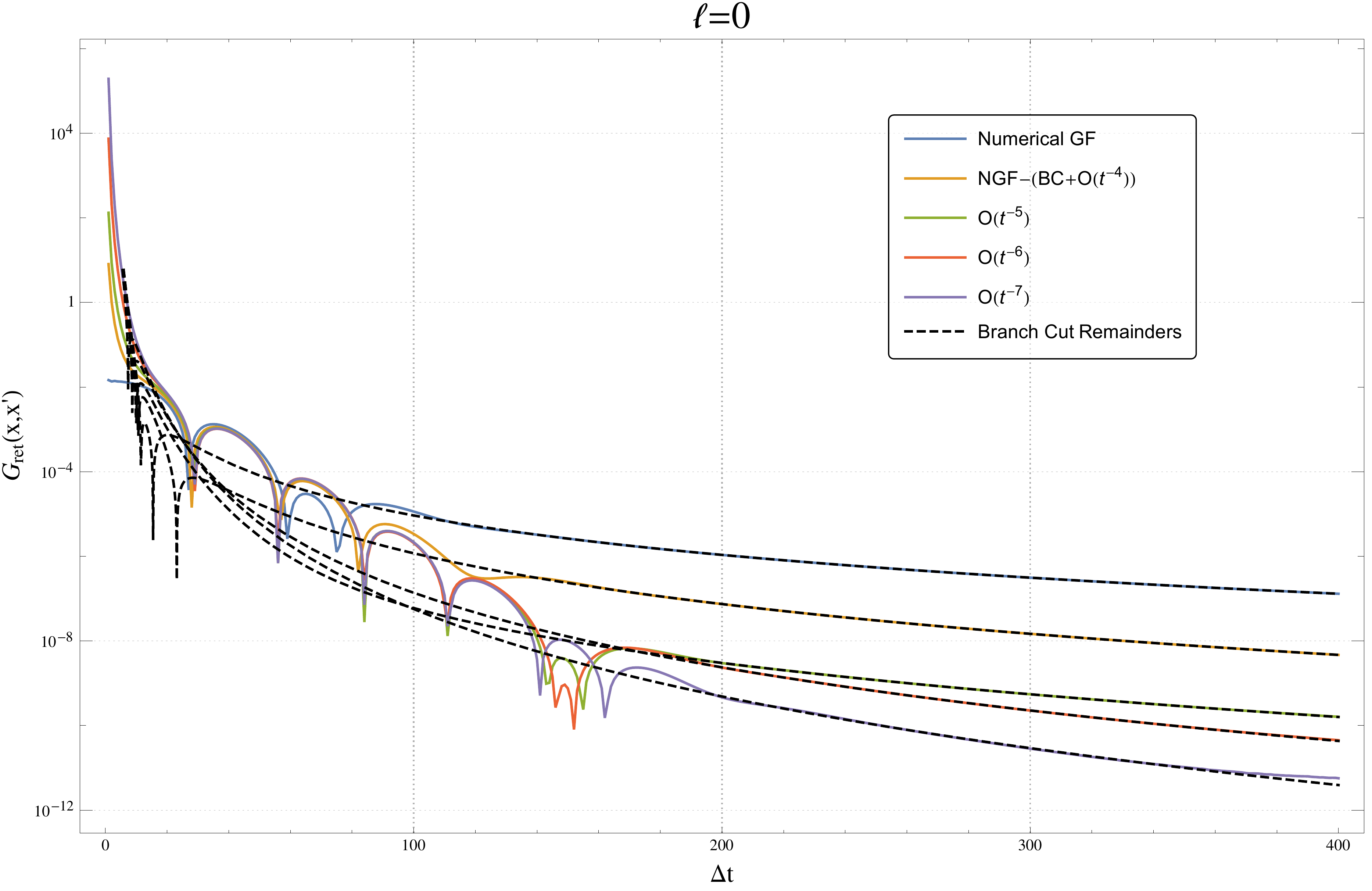}
\caption{
Scalar Green function mode $\ell=0$ as a function of time in Kerr space-time with  $a=6M/10$.
The points lie on an equatorial circular  geodesic at radius $r_0=\frac{9 \sqrt{11}}{5}M$.
The blue curve is the Green function mode obtained using a numerical real-frequency Fourier integral, which is an `exact' calculation.
The other coloured curves correspond to substracting the analytic late-time Green function up to various orders for late times from the previous real-frequency calculation.
The dashed curves are the corresponding analytic, late-time Green function remainders.
We see excellent agreement for late times $t>200M$.}
\label{fig:realfreqplots ell=0}
\end{figure}

\begin{figure}[htb!]
\includegraphics[scale=.4]{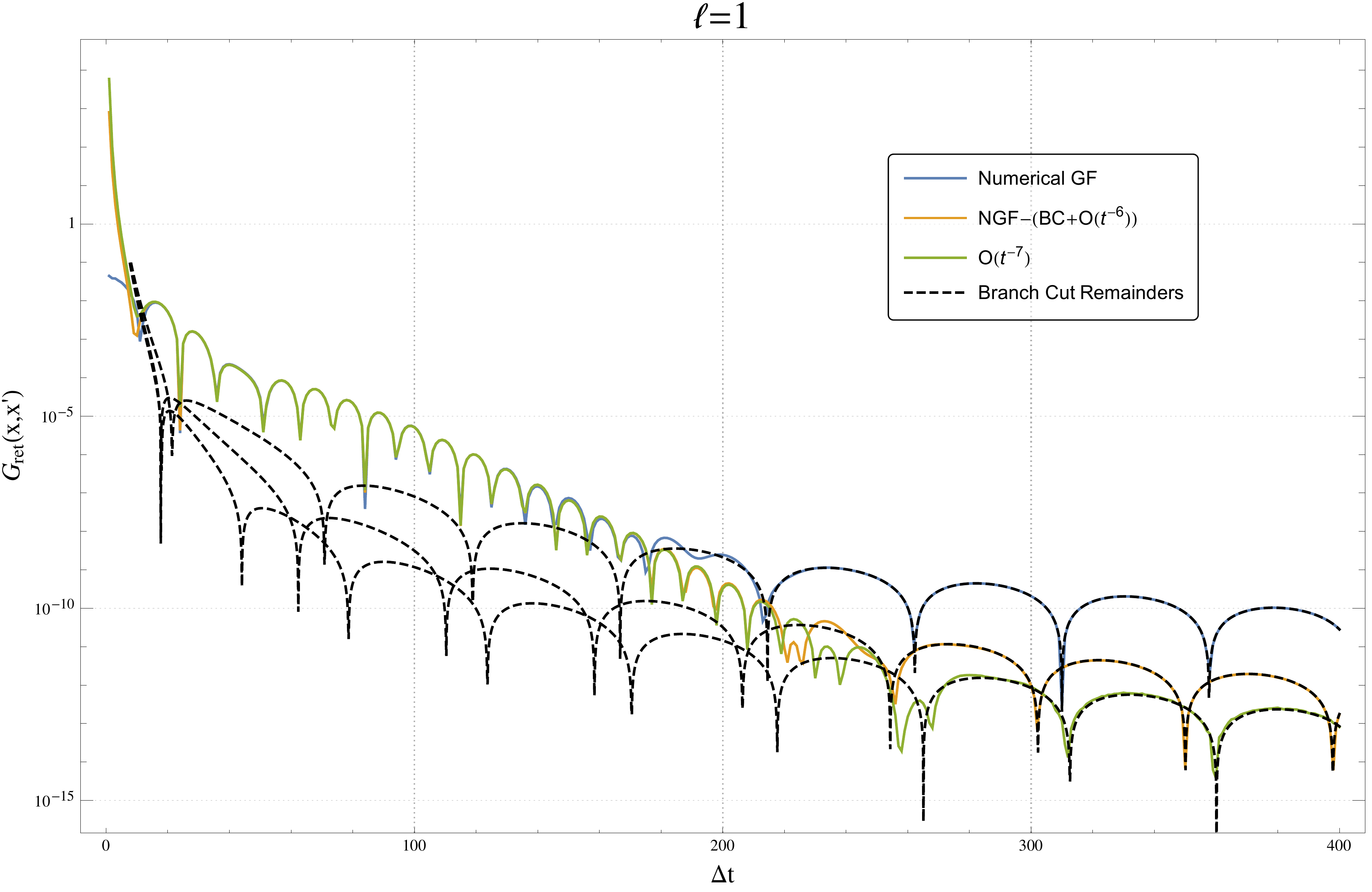}
\caption{
Same as Fig.\ref{fig:realfreqplots ell=0} but for the mode $\ell=1$.}
\label{fig:realfreqplots ell=1}
\end{figure}

For comparison purposes, we choose the values used in the analytical branch cut calculation above, namely, $a=6/10M$, $\varphi=\Omega t$ and  $r_0=\frac{9 \sqrt{11}}{5}M$.
We plot the comparisons in Figs.~\ref{fig:realfreqplots ell=0} and \ref{fig:realfreqplots ell=1}.
We find excellent agreement at late times $\Delta t >200M$ between the analytical branch cut calculation
and the `exact' numerical calculation.

\newpage

\section{Discussion}\label{sec:disc}

In this paper we have developed the analytical MST formalism for obtaining the branch cut modes of the  Green function for general integer-spin on Kerr space-time.
We then applied this formalism to derive an  expansion to five orders for small-frequency of these modes in the case of a scalar field.
This expansion can readily be used to obtain a high-order expansion of the full scalar Green function at late times for arbitrary values of the spatial coodinates.
The leading order of the expansion, for a given spheroidal $\ell$-mode, is of order  $t^{-2\ell-3}$, in agreement with previous results in the literature.
The higher orders, to the best of our knowledge had not been previously obtained.
In particular, these higher orders show that a logarithmic behaviour starts appearing at order $t^{-2\ell-5}\ln t$.

A high-order late-time expansion of the Green function can be valuable for the calculation of self-forces as carried out in~\cite{CDOW13} in the case of Schwarzschild space-time.
The next step is to obtain a similar high-order expansion for a field of higher spin.
As we have developed the MST formalism for the branch cut modes for general spin, we have already
laid much of the groundwork for extending the calculation to electromagnetic and gravitatonal fields. 
The gravitational case  is of particular interest given the recent detections of gravitational waves from black hole inspirals by the
Laser Interferometer Gravitational-Wave Observatory~\cite{PhysRevLett.116.061102,Abbott:2016nmj},
in which the ringdown stage was observed.


\begin{acknowledgments}
We are thankful to Barry Wardell for useful discussions.
M.C. acknowledges partial financial support by CNPq (Brazil), process number 308556/2014-3. 
C.K. was supported by the Programme for Research in Third Level
Institutions (PRTLI) Cycle 5 and co-funded under
the European Regional Development Fund.
\end{acknowledgments}


\appendix
\section{Angular branch cuts}
\label{sec:AngularCuts}

It has been shown (see~\cite{BONGK:2004} and references therein) that the eigenvalues of the spheroidal wave equation possess square root branch cuts, typically chosen to be emanating radially outwards from points where two eigenvalues coalesce, and the eigenvalue problem is found to have a double root. Along these `angular branch cuts', the two spheroidal eigenvalues are found to be analytic continuations of each other. 
To understand this better, let a superscript $(1)$ on a quantity that depends on the spheroidal eigenvalue/eigenfunction denote the quantity being evaluated on a Riemann sheet which we take to be the principal value of a spheroidal eigenvalue/eigenfunction. Similarly, a superscript $(2)$ will denote the analytic continuation of the eigenvalue/eigenfunction onto an `upper' Riemann sheet. (e.g. if $f(\omega)=\ln{\omega}$, then $f^{(1)}=\ln{\omega}$ and $f^{(2)}=\ln{\omega}+2 \pi i$). Then, in the vicinity of a branch cut with branch point having $ \lambda_{\ell,m}=\lambda_{\ell+2,m}$\MC{This equality seems to come out of the blue}, we have
\begin{align}
&\lambda_{\ell,m}^{(1)}=\lambda_{\ell+2,m}^{(2)},\text{ and}  \label{Riemannsheet1}\\
&\lambda_{\ell+2,m}^{(1)}=\lambda_{\ell,m}^{(2)},\text{  } m<\ell \label{Riemannsheet2}
\end{align} 
where the principle value is chosen so that the eigenvalue is continuous on a radial contour to the origin. Note that all branch cuts occur \MC{? don't they occur no matter what the $\ell$ and $m$ is? do you mean `pair up'?} between eigenvalues where the $\ell$-values differ by 2, and the $m$-values coincide. The branch cuts in the eigenvalue will then manifest as branch cuts in the radial and angular functions, and as such in the Green function modes. 

We now consider the case of the branch cut contribution to $G$ from the discontinuity shared by the $\ell=\ell'$ and $\ell=\ell'+2$ eigenvalues. We let the branch point be $c$, with a phase $\alpha$,  and the branch cut run along $a\omega=\zeta e^{i \alpha}$, with $\zeta: |c|\rightarrow\infty$. 
For notational convenience, in this appendix we ommit any $s$, $m$ or $\omega$ indices on quantities and the argument of any function will refer to the phase of the frequency 
where it is being evaluated.
Then
the contribution to the Green function from the discontinuity of its modes across this branch cut is
\begin{align*}
\Disc G&=2 e^{i \alpha}\sum_{\ell, m}\int_{|c|}^{\infty}e^{-i \omega t+i m\phi}\Disc (G_{\ell} S_{\ell} S^*_{\ell}) \text{d}\zeta\\
&=2 e^{i \alpha}\sum_{m}\int_{|c|}^{\infty}e^{-i \omega t+i m\phi}\left(\Disc (G_{\ell'} S_{\ell'} S^*_{\ell'}) +\Disc (G_{\ell'+2} S_{\ell'+2} S^*_{\ell'+2}) \right) \text{d}\zeta \\
&=2 e^{i \alpha}\sum_{m}\int_{|c|}^{\infty} e^{-i \omega t+i m\phi}\Disc I \text{d}\zeta,
\end{align*}
\MC{can we take $m=-\ell' \to \ell'$ in the sum? Note there's a summand with $\ell=\ell'+2$, but I think that for the cut at play there the $\ell=\ell'+1$,$\ell'+2$ don't contribute, right?}
where the last equality is just the definition of $\Disc I$.
Note that the $\ell \neq \ell',\ell'+2$ do not have a disconuity across the cut that we are considering here and so they do not contribute to the $Disc G$ as defined above.
Keeping the above identities \eqref{Riemannsheet1}-\eqref{Riemannsheet2} 
 in mind, we can write
\begin{align}
\Disc (G_{\ell'} S_{\ell'} S^*_{\ell'})&=\lim_{\rho\rightarrow 0^+}\left(G_{\ell'}^{(1)}(\theta+\rho) S_{\ell'}^{(1)}(\theta+\rho) S^{(1)*}_{\ell'}(\theta+\rho) -G_{\ell'}^{(1)}(\theta-\rho) S_{\ell'}^{(1)}(\theta-\rho) S^{(1)*}_{\ell'}(\theta-\rho)\right) \\
&=\lim_{\rho\rightarrow 0^+}\left(G_{\ell'}^{(1)}(\theta+\rho) S_{\ell'}^{(1)}(\theta+\rho) S^{(1)*}_{\ell'}(\theta+\rho) -G_{\ell'+2}^{(2)}(\theta-\rho) S_{\ell'+2}^{(2)}(\theta-\rho) S^{(2)*}_{\ell'+2}(\theta-\rho)\right),
\nonumber
\end{align}
and likewise for $\Disc (G_{\ell'+2} S_{\ell'+2} S^*_{\ell'+2})$. Putting these together gives 
\begin{align}
\Disc I &=\lim_{\rho\rightarrow 0^+}\left({G_{\ell'}^{(1)}(\theta+\rho) S_{\ell'}^{(1)}(\theta+\rho) S^{(1)*}_{\ell'}(\theta+\rho)} - {G_{\ell'+2}^{(2)}(\theta-\rho) S_{\ell'+2}^{(2)}(\theta-\rho) S^{(2)*}_{\ell'+2}(\theta-\rho)}\right) \nonumber \\
&\hphantom{==}+\lim_{\rho\rightarrow 0^+}\left({G_{\ell'+2}^{(1)}(\theta+\rho) S_{\ell'+2}^{(1)}(\theta+\rho) S^{(1)*}_{\ell'+2}(\theta+\rho)} -{G_{\ell'}^{(2)}(\theta-\rho) S_{\ell'}^{(2)}(\theta-\rho) S^{(2)*}_{\ell'}(\theta-\rho)}\right) \\
&=\lim_{\rho\rightarrow 0^+}\left({G_{\ell'}^{(1)}(\theta+\rho) S_{\ell'}^{(1)}(\theta+\rho) S^{(1)*}_{\ell'}(\theta+\rho)} -{G_{\ell'}^{(2)}(\theta-\rho) S_{\ell'}^{(2)}(\theta-\rho) S^{(2)*}_{\ell'}(\theta-\rho)}\right. \nonumber\\
&\hphantom{==}+\left.{G_{\ell'+2}^{(1)}(\theta+\rho) S_{\ell'+2}^{(1)}(\theta+\rho) S^{(1)*}_{\ell'+2}(\theta+\rho)} -{G_{\ell'+2}^{(2)}(\theta-\rho) S_{\ell'+2}^{(2)}(\theta-\rho) S^{(2)*}_{\ell'+2}(\theta-\rho)}\right)
\nonumber \\
&=0,
\nonumber
\end{align}
since everything here is, by definition, continuous going from one Riemann sheet to the other. 

It is worth noting here that while 
there will be no overall contribution from the angular branch cuts, there is nothing to say that we would not perhaps see some 
artifacts of them for a given $\ell$-mode, since it is the sum over $\ell$'s which is giving  the cancellations. 



\section{Series coefficients and renormalized angular momentum} 
\label{sec:MSTexpansions}

In this appendix we give the explicit low-frequency behaviour of the MST series renormalised angular momentum  $\RAM$ and series coefficients $a_n$ for a scalar field in Kerr space-time. We give expansions to order $\epsilon^4$ with the next term expected at $O(\epsilon^5)$ unless otherwise indicated.

\subsection{$\ell\geq 4$}
\label{sec:lgt4}
\begin{subequations}
\label{eq:lbigm0}
\begin{align} 
\RAM&=\ell -\frac{\left(15 \ell^2+15 \ell-11\right) }{2 (2
   \ell-1) (2 \ell+1) (2 \ell+3)}\epsilon ^2+ \frac{\left(5 \ell^2+5 \ell-3\right) m q }{\ell (\ell+1) (2 \ell-1) (2
   \ell+1) (2 \ell+3)}\epsilon ^3+\nonumber\\
&\quad\Bigl[\frac{(- 18480 \ell^{10}- 92400 \ell^9- 79800 \ell^8+ 235200 \ell^7+ 
 382305 \ell^6- 64365 \ell^5- 278260 \ell^4+ 9955 \ell^3+ 73892 \ell^2- 8733 \ell -3240 )}{8 \ell (\ell+1) (2 \ell-3) (2 \ell -1)^3 (2\ell+1)^3 (2\ell+3)^3 (2\ell+5)}+\nonumber\\
&\qquad\frac{   ( 65 \ell^4+ 130 \ell^3- 
   211 \ell^2- 276 \ell+126   ) q^2}{4 (2\ell -3) (2\ell -1)^2 (2\ell+1) (2\ell+3)^2 (2\ell+5)}-\frac{3  (85 \ell^4+ 
   170 \ell^3- 278 \ell^2- 363\ell+180    ) m^2 q^2}{4 \ell (\ell+1) (2 \ell -3) (2 \ell -1)^2 (2\ell+1) (2\ell+3)^2 (2\ell+5)}\Bigr]\epsilon^4\\
a_{4}&=\frac{(\ell+3) (\ell+ 4)}{96 (2\ell+1) (2\ell+3)^2 (2\ell+4)^2 (2\ell+7)} \Bigl[ (\ell+1) (
     \ell+2) (\ell+3) (\ell+4) (1 - q^2)^2- (35 + 30 \ell+6 \ell^2) m^2 q^2 (1 - q^2) +m^4 q^4 +\nonumber\\
&\qquad 
   2  (2\ell+5)  \left( (5 + 5 \ell+\ell^2) (1 - q^2)-m^2 q^2 \right) i m q\kappa\Bigr]\epsilon^4\\
a_{3}&=\frac{(\ell+3)}{24 (2 \ell+1) (2 \ell+3)^2 (2 \ell+5)} \Bigl[  \left( \left((3 \ell^2+12\ell+11) (1 - q^2) -m^2 q^2 \right) m q +\right.\nonumber\\
&\qquad\left.
\left(- (\ell+1) (\ell+2) (\ell+3) (1 - q^2)  +  3  (\ell+2) m^2 q^2 \right)i\kappa\right)\epsilon^3 +
\left(3 m^2 q^2 - (11 + 12 \ell+3 \ell^2) (1 - q^2)  - 
  6 (\ell+2) i m q \kappa\right)\epsilon^4\Bigr]\\
a_{2}&=\frac{(\ell+2)}{4 (2\ell+1) (2\ell+3)^2}\Bigl[\left(m^2 q^2 - (\ell+1) (\ell+2) (1 - q^2)  - (2\ell+3) i m q \kappa\right)\epsilon^2 +  \left(-2 m q +(2\ell+3) i \kappa\right)\epsilon^3\Bigr] +\nonumber\\
&\qquad 
\frac{1}{48 (\ell+1)^2 (\ell+2)^2 (-2 \ell+1) (2 \ell+1)^3 (2 \ell+3)^4 (2 \ell+7)}\Bigl[(\ell+1)^2 (\ell+2)^2 (2 \ell+1)^2 (129 + 150 \ell+60 \ell^2 + 
    8 \ell^3) m^4 q^4 +\nonumber\\
&\qquad  (\ell+2) m^2 q^2 (-20034 - 69453 \ell - 39741 \ell^2 + 
    152557 \ell^3 + 318923 \ell^4 + 271192 \ell^5 + 118400 \ell^6 + 25240 \ell^7 +\nonumber\\
&\qquad + 
    1712 \ell^8 - 96 \ell^9 + (546  + 4875 \ell+ 18471 \ell^2  + 
    38963 \ell^3  + 50443 \ell^4  + 41582 \ell^5  + 21832 \ell^6  +\nonumber\\
&\qquad  
    7040 \ell^7  + 1264 \ell^8  + 96 \ell^9) q^2) - (\ell+1) (2 + 
    \ell)^2 (-9690 - 25755 \ell+1608 \ell^2 + 86706 \ell^3 + 141666 \ell^4 +\nonumber\\
&\qquad 
    111967 \ell^5 + 51458 \ell^6 + 14800 \ell^7 + 2864 \ell^8 + 400 \ell^9 + 
    32 \ell^{10} + 11232 q^2 + 33642 \ell q^2 +  7338 \ell^2 q^2 - 
    105372 \ell^3 q^2 \nonumber\\
&\qquad- 201780 \ell^4 q^2 - 179840 \ell^5 q^2 - 
    91708 \ell^6 q^2 - 28520 \ell^7 q^2 - 5728 \ell^8 q^2 - 800 \ell^9 q^2 - 
    64 \ell^{10} q^2 - 30 q^4 - 147 \ell q^4 \nonumber\\
&\qquad+ 114 \ell^2 q^4 + 2346 \ell^3 q^4 + 
    7170 \ell^4 q^4 + 11233 \ell^5 q^4 + 10490 \ell^6 q^4 + 6040 \ell^7 q^4 + 
    2096 \ell^8 q^4 + 400 \ell^9 q^4 + 32 \ell^{10} q^4)\nonumber\\
&\qquad +(- 
  (\ell+1)^2 (\ell+2)^2 (2 \ell+1)^2 (2 \ell+3) (129 + 150 \ell+60 \ell^2 + 
    8 \ell^3) m^2 q^2 \nonumber\\
&\qquad+ 
  (\ell+2) (2 \ell+3)  (15822 + 50229 \ell+13368 \ell^2 - 152742 \ell^3 - 
    289536 \ell^4 - 247747 \ell^5 - 116150 \ell^6 - 30400 \ell^7 \nonumber\\
&\qquad- 4352 \ell^8 - 
    400 \ell^9 - 32 \ell^{10}+( - 30  - 147 \ell+ 114 \ell^2  + 
    2346 \ell^3  + 7170 \ell^4  + 11233 \ell^5  + 10490 \ell^6  \nonumber\\
&\qquad+ 
    6040 \ell^7  + 2096 \ell^8  + 400 \ell^9  + 32 \ell^{10}) q^2)) i m q \kappa\Bigr]\epsilon^4\\
a_{1}&=  \frac{1}{2 (2\ell+1)}\left(-m q+i \kappa  (\ell+1)\right)\epsilon+ \frac{1}{2 (2 \ell+1)}\epsilon ^2\nonumber\\
&\qquad+
\frac{1}{8 (\ell+1)^2 (-2 \ell+1) (2 \ell+1)^3 (2 \ell+3)^2 (2 \ell+5)}\Bigl[ 
 m q (841 + 1800 \ell - 1346 \ell^2 - 7157 \ell^3 - 7679 \ell^4 - 3485 \ell^5 - 
    674 \ell^6 \nonumber\\
&\qquad- 60 \ell^7 - 8 \ell^8 - q^2 - 4 \ell q^2 + 10 \ell^2 q^2 + 
    89 \ell^3 q^2 + 219 \ell^4 q^2 + 269 \ell^5 q^2 + 178 \ell^6 q^2 + 
    60 \ell^7 q^2 + 8 \ell^8 q^2)\nonumber\\
&\qquad -(\ell+1)^2 (2 \ell+1)^2 (11 + 9 \ell+2 \ell^2) m^3 q^3 + 
 (\ell+1)^3 (2 \ell+1)^2 (11 + 9 \ell+2 \ell^2) m^2 q^2 i\kappa \nonumber\\
&\qquad- 
  (\ell+1) (511 + 908 \ell - 870 \ell^2 - 3835 \ell^3 - 4075 \ell^4 - 1925 \ell^5 - 
    434 \ell^6 - 60 \ell^7 - 8 \ell^8 - q^2 - 4 \ell q^2 + 10 \ell^2 q^2 \nonumber\\
&\qquad+ 
    89 \ell^3 q^2 + 219 \ell^4 q^2 + 269 \ell^5 q^2 + 178 \ell^6 q^2 + 
    60 \ell^7 q^2 + 8 \ell^8 q^2) i \kappa\Bigr]\epsilon^3 \nonumber\\
&\qquad+\frac{1}{8 \ell (\ell+2) (\ell+1)^2 (-2 \ell+1) (2 \ell+1)^3 (2 \ell+3)^2 (2 \ell+5)}\Bigl[(-900 - 1254 \ell+2979 \ell^2 + 8971 \ell^3 + 9163 \ell^4 + 4667 \ell^5  \nonumber\\
&\qquad+ 
    1310 \ell^6 + 228 \ell^7 + 24 \ell^8) m^2 q^2 - 
 \ell (\ell+2) (841 + 1800 \ell - 1346 \ell^2 - 7157 \ell^3 - 7679 \ell^4 - 
    3485 \ell^5 - 674 \ell^6 - 60 \ell^7  \nonumber\\
&\qquad- 8 \ell^8 - q^2 - 4 \ell q^2 + 
    10 \ell^2 q^2 + 89 \ell^3 q^2 + 219 \ell^4 q^2 + 269 \ell^5 q^2 + 
    178 \ell^6 q^2 + 60 \ell^7 q^2 + 8 \ell^8 q^2)  \nonumber\\
&\qquad- 
 2 (\ell+1) (-270 - 296 \ell+747 \ell^2 + 2311 \ell^3 + 2569 \ell^4 + 
    1425 \ell^5 + 426 \ell^6 + 76 \ell^7 + 8 \ell^8) i m q \kappa\Bigr]\epsilon^4
\end{align}
\end{subequations}
while $a_{-i}(\ell)=a_i(-\ell -1)$ for $i=1\dots 4$. As $a_{-1}$ is of particular importance to our argument below, we give it explicitly
\begin{align}
a_{-1}&=  \frac{1}{2 (2\ell+1)}\left(m q+i \kappa \ell \right)\epsilon- \frac{1}{2 (2\ell+1)}\epsilon ^2\nonumber\\
&\qquad+\frac{1}{8 \ell^2 (\ell-1) (\ell+1 ) (2\ell-3 ) (2 \ell-1)^2 (2\ell+1)^3 (2\ell+3)}\Bigl[- 
 m q (36 - 173 \ell^2 - 537 \ell^3 + 1176 \ell^4 + 253 \ell^5 - 478 \ell^6 - 4 \ell^7 - 
    8 \ell^8 \nonumber\\
&\qquad+ \ell^2 q^2 + 5 \ell^3 q^2 + 4 \ell^4 q^2 - 13 \ell^5 q^2 - 
    18 \ell^6 q^2 + 4 \ell^7 q^2 + 8 \ell^8 q^2)+ \ell^2 (2\ell+1)^2 (2 \ell^2- 5 \ell+4 ) m^3 q^3  \nonumber\\
&\qquad+ 
  \ell^3 (2 \ell+1)^2 (4 - 5 \ell+2 \ell^2) m^2 q^2 i\kappa - 
 \ell (36 - 66 \ell - 39 \ell^2 - 243 \ell^3 + 580 \ell^4 + 133 \ell^5 - 238 \ell^6 - 
    4 \ell^7 - 8 \ell^8 \nonumber\\
&\qquad+ \ell^2 q^2 + 5 \ell^3 q^2 + 4 \ell^4 q^2 - 13 \ell^5 q^2 - 
    18 \ell^6 q^2 + 4 \ell^7 q^2 + 8 \ell^8 q^2) i \kappa\Bigr]\epsilon ^3\nonumber\\
&\qquad+
\frac{1}{8 \ell^2 (\ell-1) (\ell+1) (2 \ell-3) (2 \ell-1)^2 (2\ell+1)^3 (2\ell+3)}\Bigl[ (\ell-1) (\ell+1) (36 - 173 \ell^2 - 
    537 \ell^3 + 1176 \ell^4 + 253 \ell^5 - 478 \ell^6 \nonumber\\
&\qquad- 4 \ell^7 - 8 \ell^8 + 
    \ell^2 q^2 + 5 \ell^3 q^2 + 4 \ell^4 q^2 - 13 \ell^5 q^2 - 18 \ell^6 q^2 + 
    4 \ell^7 q^2 + 8 \ell^8 q^2) \nonumber\\
&\qquad-(-36 + 72 \ell - 92 \ell^2 + 575 \ell^3 - 822 \ell^4 - 251 \ell^5 + 386 \ell^6 - 
     36 \ell^7 + 24 \ell^8) m^2 q^2 \nonumber\\
&\qquad- 
 2 i \ell (-36 + 96 \ell - 4 \ell^2 + 23 \ell^3 - 266 \ell^4 - 17 \ell^5 + 118 \ell^6 - 
    12 \ell^7 + 8 \ell^8) m q \kappa\Bigr]\epsilon ^4.
\end{align}

An examination of these expressions shows that they contain terms that are singular when $\ell=0$ and $\ell=1$. For example,
consider the coefficient of $\epsilon^3$ in $a_{-1}(0)=a_{1}(-1)$ and $\epsilon^4$ in $a_{-1}(1)=a_1(-2)$.
Indeed, from the structure of the MST recurrence relation one can see that for any given $\ell$ there are singular terms 
in $a_{-1}(\ell)$ at order $\ell+3$. In addition, propagation of these
 singular terms leads to anomalous behaviour even in terms that are finite at order $\ell+1$.
Note that comments here relate to the solution with $\nu =\ell+O(\epsilon^2)$, the corresponding singularities for $\nu = -\ell-1+O(\epsilon^2)$ first appear in $a_1$.

In the following subsections we give the behaviour for the low-$\ell$ terms.  As the expansions for $\ell=0$ and $\ell=1$ 
contain singular terms we give them in full, while for $\ell=2$ and $\ell=3$ we simply give the anomalous contributions.

\subsection{$\ell=0$, $m=0$}
\begin{subequations}
\label{eq:l0m0}
\begin{align} 
\RAM&=
-\frac{7}{6} \epsilon ^2 +  \left(-\frac{9449}{7560}+\frac{3 q^2}{35}\right) \epsilon ^4 +O(\epsilon ^6)\\
a_{4}&=\frac{1}{525} \left(1-q^2\right)^2 \epsilon ^4\\
a_{3}&=-\frac{1}{60} i
    \left( 1- q^2 \right)\kappa \epsilon ^3 -\frac{11}{360} \left(1-q^2\right) \epsilon ^4\\
a_{2}&=-\frac{1}{9} \left(1-q^2\right) \epsilon ^2+\frac{1}{6} i \kappa  \epsilon^3-\frac{1}{4536}
\left(943-1200   q^2+5 q^4\right) \epsilon ^4\\
a_{1}&=\frac{i \kappa   }{2}\epsilon+\frac{1}{2} \epsilon ^2+\frac{1}{360} i  \left(331-  q^2 
   \right)\kappa \epsilon ^3 +\frac{1}{360} \left(541-q^2\right) \epsilon
   ^4\\
a_{-1}&=-\frac{2}{9}-\frac{7  }{27}i \kappa  \epsilon-\frac{1}{5670}\left(3182+51 q^2\right)
   \epsilon ^2 -\frac{1}{34020} \left(31723-291 q^2 
   \right)i \kappa \epsilon ^3 \nonumber\\
&\qquad-\frac{1}{50009400}\left(73956143-583931 q^2+25673 q^4\right) \epsilon
   ^4\\
a_{-2}&=-\frac{1 }{9} i \kappa  \epsilon +\frac{1}{54}\left(1-7 q^2\right) \epsilon
   ^2-\frac{1}{11340} \left(2559  +44  
   q^2\right)i \kappa \epsilon ^3 +\frac{1}{34020}\left(9679-8838 q^2+170 q^4\right) \epsilon ^4\\
a_{-3}&=\frac{2}{81} \left(1-q^2\right) \epsilon ^2-\frac{1}{243} \left(2  +7  
   q^2\right)i \kappa \epsilon ^3 +\frac{1}{102060}\left(3729-4912 q^2-77
   q^4\right) \epsilon ^4\\
a_{-4}&=\frac{1}{270}
   \left( 1- q^2\right) i \kappa \epsilon ^3+\frac{1}{1620}\left(4+3 q^2-7 q^4\right) \epsilon ^4\\
a_{-5}&=-\frac{2}{4725} \left(1-q^2\right)^2 \epsilon ^4
\end{align}
\end{subequations}
\goodbreak 

\subsection{$\ell=1$, $m=0$}
\begin{subequations}
\label{eq:l1m0}
\begin{align} 
\RAM&=1-\frac{19 }{30}\epsilon ^2+\left(-\frac{1325203}{3591000}+\frac{117 q^2}{3325}\right) \epsilon  ^4 +O(\epsilon ^6)\\
a_{4}&=\frac{1}{1323}\left(1-q^2\right)^2 \epsilon ^4\\
a_{3}&=-\frac{4}{525}  \left( 1- q^2 \right) i\kappa
   \epsilon ^3-\frac{13}{1575} \left(1-q^2\right) \epsilon ^4\\
a_{2}&=-\frac{3}{50} \left(1-q^2\right) \epsilon ^2+\frac{1}{20} i \kappa 
   \epsilon ^3-\frac{1}{67500}\left(1741-2600 q^2+184
   q^4\right) \epsilon ^4\\
a_{1}&=\frac{1
   }{3}i \kappa  \epsilon+\frac{1}{6}\epsilon ^2+\frac{1}{18900} \left(2447-207   q^2 
   \right) i \kappa \epsilon ^3 +\frac{1}{37800}\left(4442-207 q^2\right) \epsilon
   ^4\\
a_{-1}&=\frac{1
   }{6}i \kappa  \epsilon-\frac{1}{6}\epsilon ^2+\frac{1}{20520} \left(169-171   q^2 
   \right)i\kappa
   \epsilon ^3-\frac{1}{20520}\left(2335-171 q^2\right) \epsilon ^4\\
a_{-2}&=\frac{5}{4332}
    \left(109 +5   q^2\right)i \kappa\epsilon ^3-\frac{1}{25992}\left(5341-1826 q^2-95 q^4\right) \epsilon ^4\\
a_{-3}&=-\frac{25}{722} \left(1-q^2\right) \epsilon ^2-\frac{5}{228}  \left( 1- q^2
   \right)i \kappa \epsilon ^3 -\frac{1}{65681784}\left(10140442-7387693 q^2-478449
   q^4\right) \epsilon ^4\\
a_{-4}&=-\frac{25}{2166} \left(1-  q^2\right) i \kappa
   \epsilon ^3+\frac{5}{12996} \left(4-23 q^2+19 q^4\right) \epsilon ^4\\
a_{-5}&=\frac{3}{1444} \left(1-q^2\right)^2 \epsilon ^4
\end{align}
\end{subequations}

\subsection{$\ell=1$, $m=\pm 1$}
\begin{subequations}
\label{eq:l1mn0}
\begin{align} 
\RAM&=1-\frac{19 }{30}\epsilon ^2+\frac{18  }{95} m q \epsilon ^3-\left(\frac{1325203}{3591000}-\frac{17947 q^2}{3600975}\right) \epsilon
   ^4\\
a_{4}&=\frac{1}{158760}\left(\left(120-311 q^2+192 q^4\right)+\left(154-168  q^2\right) i m
   \kappa  q\right)\epsilon ^4 \\
a_{3}&=\frac{1}{3150}  \left(
   (26-27 q^2) m q+(-24+33  q^2) i \kappa\right)\epsilon ^3 + \frac{1}{3150} \left((-26+29 q^2)-18 i \kappa
    m q\right) \epsilon ^4 \\
a_{2}&=\frac{1}{100} 
   \left((-6+7 q^2)-5 i \kappa  m q\right)\epsilon ^2 +\frac{1}{100} (-2 m q+5 i \kappa )\epsilon ^3 +\nonumber\\
&\qquad \frac{1}{810000} \left((-20892+38225
   q^2-147 q^4)+5(-7055+21   q^2) i \kappa  m q\right)\epsilon ^4\\
a_{1}&=
  \frac{1}{6} \left(-m q+2 i \kappa\right)\epsilon +\frac{1}{6}\epsilon
   ^2+\frac{1}{37800} \left(
  ( -4442 +9 q^2)m q+2(2447 - 9 q^2)i \kappa\right)\epsilon ^3+\nonumber\\
&\qquad \frac{1}{1436400} \left(168796+57021
   q^2-69366 i \kappa  m q\right)\epsilon ^4\\
a_{-1}&=\frac{1}{6}   (m q+i
   \kappa )\epsilon+\frac{1}{114}  \left(-19+15
   q^2+15 i \kappa  m q\right)\epsilon ^2+\frac{1}{7407720} \left((-131765 +201888
   q^2)m q+(61009+201888 
   q^2) i \kappa \right)\epsilon ^3+\nonumber\\
&\qquad\frac{1}{18719308440}\left((-2130096745+4509021765
   q^2-340902270 q^4)+(4050386427-340902270 q^2) i \kappa  m q\right)\epsilon ^4 \\
a_{-2}&=-\frac{545}{4332} 
   \left(q^2+i \kappa  m q\right)\epsilon ^2-\frac{5 }{1563852}
 \left( ( -78698 +16142 q^2) m q+( -39349 +16142 
   q^2)i \kappa\right)\epsilon ^3+\nonumber\\
&\qquad\frac{1}{711333720720} \left((-146169336810-195368785331
   q^2+2456535750 q^4 )+(-175402629821+2456535750 q^2) i \kappa  m q\right)\epsilon ^4\\
a_{-3}&=\frac{25  }{722}m q \epsilon -\frac{25}{260642}
   \left(481 q^2+361\right) \epsilon ^2+\left(\frac{1}{11855562012} \left(2377318931-488743146
   q^2\right)m q-\frac{5}{228}  \left(1
   +2   q^2\right)i \kappa\right)\epsilon ^3 \nonumber\\
&\qquad -\frac{
   \left((1321512541882+563785252722  q^2+114839191503  q^4)+(138314890140  q^2-1057120946070)  i\kappa  m
   q)
   \right)}{8559715772664}	\epsilon ^4\\
a_{-4}&=-\frac{25}{4332} q  (q-2 i
   \kappa  m)\epsilon ^2+\frac{25}{1563852}
   \left( \left(722+481 q^2\right)m q-2  \left(361+481
   q^2\right) i \kappa\right) \epsilon ^3+\nonumber\\
&\qquad \frac{1}{142266744144} \left((218939280-3914221559
   q^2-1101263979 q^4)+(8760451258 -917356782q^2)
   i \kappa  m q\right)\epsilon ^4\\
a_{-5}&=\frac{1}{2888} q \left(m
   \left(-6+7 q^2\right)-5 i \kappa  q\right) \epsilon ^3+\frac{1}{1042568}
   \left((2166-363
   q^2-3367 q^4)+(3610+2405 q^2 ) i \kappa  m q\right)\epsilon ^4\\
a_{-6}&=\frac{1}{90972}  \left((26 -27 q^2)q^2+(-24+33  q^2) i
   \kappa  mq \right)\epsilon ^4
\end{align}
\end{subequations}

For $\ell=2$ and $\ell=3$ the $l\geq4$, expansions to order $\epsilon^4$ contain only finite terms 
but certain terms do not correspond to the correct answer (obtained by setting $l$ to its appropriate 
value at the beginning of the calculation) -- this feature already occurred in Schwarzschild space-time in~\cite{Casals:Ottewill:2015}. 
In the cases given below, the terms  $\Delta a_{i}$ must be added to the
general $l$ expansions of Subsection~\ref{sec:lgt4} to obtain the correct values. In all other cases, the general $l$ 
expansions yield the correct values to this order.

\subsection{$\ell=2$, $m\neq 0$}
\begin{subequations}
\label{eq:l1mn0}
\begin{align} 
\Delta a_{-1}&=-\frac{7}{11376}  m q \left((1 -q^2) +  m^2 q^2\right)  (m q + 2 i \kappa)\epsilon^4\\
\Delta a_{-2}&=-\frac{7}{2844} m q \left(\left(m^2+2\right) q^2+3 i
   m \kappa  q-2\right) \epsilon ^3\nonumber\\
&\qquad-\frac{7}{224676} \left(9 m^2 \left(m^2+2\right) q^4+27 i m^3 \kappa 
   q^3-\left(255 m^2+158\right) q^2-474 i m \kappa  q+158\right)
   \epsilon ^4\\
\Delta a_{-3}&=\frac{7}{2844} m q \left(\left(m^2+2\right) q^2+3 i
   m \kappa  q-2\right) \epsilon ^3\nonumber\\
&\qquad+\frac{7}{224676} \left(9 m^2 \left(m^2+2\right) q^4+27 i m^3 \kappa 
   q^3-\left(255 m^2+158\right) q^2-474 i m \kappa  q+158\right)
   \epsilon ^4\\
\Delta a_{-4}&=-\frac{7}{4493520} m q \left(
\left(339-360
   m^2\right) q^2-367\right)  (m q+2 i \kappa ) \epsilon ^4 \\
\Delta a_{-5}&=-\frac{49}{449352} m q
   \left(\left(m^2-4\right) q^2+4\right) \left(\left(m^2-1\right)
   q^2+1\right) \epsilon ^3\nonumber\\
&\qquad-\frac{49}{35498808} \left(
\left(-305
   m^4+1041 m^2-316\right) q^4+\left(632-1113 m^2\right)
   q^2-316\right) \epsilon ^4\\
\Delta a_{-6}&=\frac{49}{4493520} m q \left(\left(m^2-4\right) q^2+4\right)
   \left(\left(m^2-1\right) q^2+1\right) (m q-3 i \kappa
   ) \epsilon ^4
\end{align}
\end{subequations}

\subsection{$\ell=3$, $m\neq 0$}
\begin{subequations}
\label{eq:l1mn0}
\begin{align} 
\Delta a_{-3}&=-\frac{1}{40560}m q \left(m^3 q^3+6 i m^2 \kappa  q^2+11 m
   \left(q^2-1\right) q+6 i \left(q^2-1\right) \kappa \right) \epsilon ^4 \\
\Delta a_{-4}&=\frac{1}{40560}m q  \left(m^3 q^3+6 i m^2 \kappa  q^2+11 m
   \left(q^2-1\right) q+6 i \left(q^2-1\right) \kappa \right)\epsilon ^4
\end{align}
\end{subequations}



\begin{thebibliography}{40}
\expandafter\ifx\csname natexlab\endcsname\relax\def\natexlab#1{#1}\fi
\expandafter\ifx\csname bibnamefont\endcsname\relax
  \def\bibnamefont#1{#1}\fi
\expandafter\ifx\csname bibfnamefont\endcsname\relax
  \def\bibfnamefont#1{#1}\fi
\expandafter\ifx\csname citenamefont\endcsname\relax
  \def\citenamefont#1{#1}\fi
\expandafter\ifx\csname url\endcsname\relax
  \def\url#1{\texttt{#1}}\fi
\expandafter\ifx\csname urlprefix\endcsname\relax\def\urlprefix{URL }\fi
\providecommand{\bibinfo}[2]{#2}
\providecommand{\eprint}[2][]{\url{#2}}

\bibitem[{\citenamefont{Regge and Wheeler}(1957)}]{Regge:1957td}
\bibinfo{author}{\bibfnamefont{T.}~\bibnamefont{Regge}} \bibnamefont{and}
  \bibinfo{author}{\bibfnamefont{J.~A.} \bibnamefont{Wheeler}},
  \bibinfo{journal}{Phys. Rev.} \textbf{\bibinfo{volume}{108}},
  \bibinfo{pages}{1063} (\bibinfo{year}{1957}).

\bibitem[{\citenamefont{Zerilli}(1970)}]{PhysRevD.2.2141}
\bibinfo{author}{\bibfnamefont{F.~J.} \bibnamefont{Zerilli}},
  \bibinfo{journal}{Phys. Rev. D} \textbf{\bibinfo{volume}{2}},
  \bibinfo{pages}{2141} (\bibinfo{year}{1970}),
  \urlprefix\url{http://link.aps.org/doi/10.1103/PhysRevD.2.2141}.

\bibitem[{\citenamefont{Whiting}(1989)}]{whiting1989mode}
\bibinfo{author}{\bibfnamefont{B.}~\bibnamefont{Whiting}},
  \bibinfo{journal}{Journal of Mathematical Physics}
  \textbf{\bibinfo{volume}{30}}, \bibinfo{pages}{1301} (\bibinfo{year}{1989}).

\bibitem[{\citenamefont{Hartle and Wilkins}(1974)}]{Hartle:Wilkins:1974}
\bibinfo{author}{\bibfnamefont{J.~B.} \bibnamefont{Hartle}} \bibnamefont{and}
  \bibinfo{author}{\bibfnamefont{D.~C.} \bibnamefont{Wilkins}},
  \bibinfo{journal}{Commun. Math. Phys.} \textbf{\bibinfo{volume}{38}},
  \bibinfo{pages}{47} (\bibinfo{year}{1974}).

\bibitem[{\citenamefont{Casals et~al.}(2016)\citenamefont{Casals, Gralla, and
  Zimmerman}}]{casals2016horizon}
\bibinfo{author}{\bibfnamefont{M.}~\bibnamefont{Casals}},
  \bibinfo{author}{\bibfnamefont{S.~E.} \bibnamefont{Gralla}},
  \bibnamefont{and}
  \bibinfo{author}{\bibfnamefont{P.}~\bibnamefont{Zimmerman}},
  \bibinfo{journal}{arXiv preprint arXiv:1606.08505}  (\bibinfo{year}{2016}).

\bibitem[{\citenamefont{Poisson et~al.}(2011)\citenamefont{Poisson, Pound, and
  Vega}}]{Poisson:2011nh}
\bibinfo{author}{\bibfnamefont{E.}~\bibnamefont{Poisson}},
  \bibinfo{author}{\bibfnamefont{A.}~\bibnamefont{Pound}}, \bibnamefont{and}
  \bibinfo{author}{\bibfnamefont{I.}~\bibnamefont{Vega}},
  \bibinfo{journal}{Living Rev. Rel.} \textbf{\bibinfo{volume}{14}},
  \bibinfo{pages}{7} (\bibinfo{year}{2011}), \eprint{1102.0529}.

\bibitem[{\citenamefont{Abbott et~al.}(2016{\natexlab{a}})\citenamefont{Abbott,
  Abbott, Abbott, Abernathy, Acernese, Ackley, Adams, Adams, Addesso, Adhikari
  et~al.}}]{PhysRevLett.116.061102}
\bibinfo{author}{\bibfnamefont{B.~P.} \bibnamefont{Abbott}},
  \bibinfo{author}{\bibfnamefont{R.}~\bibnamefont{Abbott}},
  \bibinfo{author}{\bibfnamefont{T.~D.} \bibnamefont{Abbott}},
  \bibinfo{author}{\bibfnamefont{M.~R.} \bibnamefont{Abernathy}},
  \bibinfo{author}{\bibfnamefont{F.}~\bibnamefont{Acernese}},
  \bibinfo{author}{\bibfnamefont{K.}~\bibnamefont{Ackley}},
  \bibinfo{author}{\bibfnamefont{C.}~\bibnamefont{Adams}},
  \bibinfo{author}{\bibfnamefont{T.}~\bibnamefont{Adams}},
  \bibinfo{author}{\bibfnamefont{P.}~\bibnamefont{Addesso}},
  \bibinfo{author}{\bibfnamefont{R.~X.} \bibnamefont{Adhikari}},
  \bibnamefont{et~al.} (\bibinfo{collaboration}{LIGO Scientific Collaboration
  and Virgo Collaboration}), \bibinfo{journal}{Phys. Rev. Lett.}
  \textbf{\bibinfo{volume}{116}}, \bibinfo{pages}{061102}
  (\bibinfo{year}{2016}{\natexlab{a}}),
  \urlprefix\url{http://link.aps.org/doi/10.1103/PhysRevLett.116.061102}.

\bibitem[{\citenamefont{Price}(1972{\natexlab{a}})}]{Price:1971fb}
\bibinfo{author}{\bibfnamefont{R.~H.} \bibnamefont{Price}},
  \bibinfo{journal}{Phys. Rev.} \textbf{\bibinfo{volume}{D5}},
  \bibinfo{pages}{2419} (\bibinfo{year}{1972}{\natexlab{a}}).

\bibitem[{\citenamefont{Price}(1972{\natexlab{b}})}]{Price:1972pw}
\bibinfo{author}{\bibfnamefont{R.~H.} \bibnamefont{Price}},
  \bibinfo{journal}{Phys. Rev.} \textbf{\bibinfo{volume}{D5}},
  \bibinfo{pages}{2439} (\bibinfo{year}{1972}{\natexlab{b}}).

\bibitem[{\citenamefont{Leaver}(1986{\natexlab{a}})}]{Leaver:1986}
\bibinfo{author}{\bibfnamefont{E.~W.} \bibnamefont{Leaver}},
  \bibinfo{journal}{Phys. Rev. D} \textbf{\bibinfo{volume}{34}},
  \bibinfo{pages}{384} (\bibinfo{year}{1986}{\natexlab{a}}).

\bibitem[{\citenamefont{Hod}(2000{\natexlab{a}})}]{PhysRevLett.84.10}
\bibinfo{author}{\bibfnamefont{S.}~\bibnamefont{Hod}}, \bibinfo{journal}{Phys.
  Rev. Lett.} \textbf{\bibinfo{volume}{84}}, \bibinfo{pages}{10}
  (\bibinfo{year}{2000}{\natexlab{a}}),
  \urlprefix\url{http://link.aps.org/doi/10.1103/PhysRevLett.84.10}.

\bibitem[{\citenamefont{Hod}(2000{\natexlab{b}})}]{hod2000mode}
\bibinfo{author}{\bibfnamefont{S.}~\bibnamefont{Hod}},
  \bibinfo{journal}{Physical Review D} \textbf{\bibinfo{volume}{61}},
  \bibinfo{pages}{064018} (\bibinfo{year}{2000}{\natexlab{b}}).

\bibitem[{\citenamefont{Barack and Ori}(1999)}]{barack1999late}
\bibinfo{author}{\bibfnamefont{L.}~\bibnamefont{Barack}} \bibnamefont{and}
  \bibinfo{author}{\bibfnamefont{A.}~\bibnamefont{Ori}},
  \bibinfo{journal}{Physical review letters} \textbf{\bibinfo{volume}{82}},
  \bibinfo{pages}{4388} (\bibinfo{year}{1999}).

\bibitem[{\citenamefont{Barack}(1999)}]{PhysRevD.61.024026}
\bibinfo{author}{\bibfnamefont{L.}~\bibnamefont{Barack}},
  \bibinfo{journal}{Phys. Rev. D} \textbf{\bibinfo{volume}{61}},
  \bibinfo{pages}{024026} (\bibinfo{year}{1999}),
  \urlprefix\url{http://link.aps.org/doi/10.1103/PhysRevD.61.024026}.

\bibitem[{\citenamefont{Aretakis}(2015)}]{Aretakis:2012ei}
\bibinfo{author}{\bibfnamefont{S.}~\bibnamefont{Aretakis}},
  \bibinfo{journal}{Adv. Theor. Math. Phys.} \textbf{\bibinfo{volume}{19}},
  \bibinfo{pages}{507} (\bibinfo{year}{2015}), \eprint{1206.6598}.

\bibitem[{\citenamefont{Aretakis}(2012)}]{aretakis2012decay}
\bibinfo{author}{\bibfnamefont{S.}~\bibnamefont{Aretakis}},
  \bibinfo{journal}{Journal of Functional Analysis}
  \textbf{\bibinfo{volume}{263}}, \bibinfo{pages}{2770} (\bibinfo{year}{2012}).

\bibitem[{\citenamefont{Zengino?lu et~al.}(2014)\citenamefont{Zengino?lu,
  Khanna, and Burko}}]{Zenginoglu:2012us}
\bibinfo{author}{\bibfnamefont{A.}~\bibnamefont{Zengino?lu}},
  \bibinfo{author}{\bibfnamefont{G.}~\bibnamefont{Khanna}}, \bibnamefont{and}
  \bibinfo{author}{\bibfnamefont{L.~M.} \bibnamefont{Burko}},
  \bibinfo{journal}{Gen. Rel. Grav.} \textbf{\bibinfo{volume}{46}},
  \bibinfo{pages}{1672} (\bibinfo{year}{2014}), \eprint{1208.5839}.

\bibitem[{\citenamefont{Le~Tiec et~al.}(2012)\citenamefont{Le~Tiec, Barausse,
  and Buonanno}}]{le2012gravitational}
\bibinfo{author}{\bibfnamefont{A.}~\bibnamefont{Le~Tiec}},
  \bibinfo{author}{\bibfnamefont{E.}~\bibnamefont{Barausse}}, \bibnamefont{and}
  \bibinfo{author}{\bibfnamefont{A.}~\bibnamefont{Buonanno}},
  \bibinfo{journal}{Physical review letters} \textbf{\bibinfo{volume}{108}},
  \bibinfo{pages}{131103} (\bibinfo{year}{2012}).

\bibitem[{\citenamefont{Casals et~al.}(2013)\citenamefont{Casals, Dolan,
  Ottewill, and Wardell}}]{CDOW13}
\bibinfo{author}{\bibfnamefont{M.}~\bibnamefont{Casals}},
  \bibinfo{author}{\bibfnamefont{S.}~\bibnamefont{Dolan}},
  \bibinfo{author}{\bibfnamefont{A.~C.} \bibnamefont{Ottewill}},
  \bibnamefont{and} \bibinfo{author}{\bibfnamefont{B.}~\bibnamefont{Wardell}},
  \bibinfo{journal}{Phys. Rev. D} \textbf{\bibinfo{volume}{88}},
  \bibinfo{pages}{044022} (\bibinfo{year}{2013}),
  \urlprefix\url{http://link.aps.org/doi/10.1103/PhysRevD.88.044022}.

\bibitem[{\citenamefont{Casals and
  Ottewill}(2012{\natexlab{a}})}]{PhysRevLett.109.111101}
\bibinfo{author}{\bibfnamefont{M.}~\bibnamefont{Casals}} \bibnamefont{and}
  \bibinfo{author}{\bibfnamefont{A.}~\bibnamefont{Ottewill}},
  \bibinfo{journal}{Phys. Rev. Lett.} \textbf{\bibinfo{volume}{109}},
  \bibinfo{pages}{111101} (\bibinfo{year}{2012}{\natexlab{a}}),
  \urlprefix\url{http://link.aps.org/doi/10.1103/PhysRevLett.109.111101}.

\bibitem[{\citenamefont{Casals and Ottewill}(2013)}]{Casals:2012ng}
\bibinfo{author}{\bibfnamefont{M.}~\bibnamefont{Casals}} \bibnamefont{and}
  \bibinfo{author}{\bibfnamefont{A.~C.} \bibnamefont{Ottewill}},
  \bibinfo{journal}{Phys.Rev.} \textbf{\bibinfo{volume}{D87}},
  \bibinfo{pages}{064010} (\bibinfo{year}{2013}), \eprint{1210.0519}.

\bibitem[{\citenamefont{Casals and
  Ottewill}(2012{\natexlab{b}})}]{Casals:2011aa}
\bibinfo{author}{\bibfnamefont{M.}~\bibnamefont{Casals}} \bibnamefont{and}
  \bibinfo{author}{\bibfnamefont{A.}~\bibnamefont{Ottewill}},
  \bibinfo{journal}{Phys.Rev.} \textbf{\bibinfo{volume}{D86}},
  \bibinfo{pages}{024021} (\bibinfo{year}{2012}{\natexlab{b}}),
  \eprint{1112.2695}.

\bibitem[{\citenamefont{Casals and Ottewill}(2015)}]{Casals:Ottewill:2015}
\bibinfo{author}{\bibfnamefont{M.}~\bibnamefont{Casals}} \bibnamefont{and}
  \bibinfo{author}{\bibfnamefont{A.}~\bibnamefont{Ottewill}},
  \bibinfo{journal}{Phys. Rev. D} \textbf{\bibinfo{volume}{92}},
  \bibinfo{pages}{124055} (\bibinfo{year}{2015}),
  \urlprefix\url{http://link.aps.org/doi/10.1103/PhysRevD.92.124055}.

\bibitem[{\citenamefont{Casals and Ottewill}(2016)}]{casals2016quasi}
\bibinfo{author}{\bibfnamefont{M.}~\bibnamefont{Casals}} \bibnamefont{and}
  \bibinfo{author}{\bibfnamefont{A.~C.} \bibnamefont{Ottewill}},
  \bibinfo{journal}{arXiv preprint arXiv:1606.03423}  (\bibinfo{year}{2016}).

\bibitem[{\citenamefont{Mano et~al.}(1996)\citenamefont{Mano, Suzuki, and
  Takasugi}}]{Mano:Suzuki:Takasugi:1996}
\bibinfo{author}{\bibfnamefont{S.}~\bibnamefont{Mano}},
  \bibinfo{author}{\bibfnamefont{H.}~\bibnamefont{Suzuki}}, \bibnamefont{and}
  \bibinfo{author}{\bibfnamefont{E.}~\bibnamefont{Takasugi}},
  \bibinfo{journal}{Prog. Theor. Phys.} \textbf{\bibinfo{volume}{95}},
  \bibinfo{pages}{1079} (\bibinfo{year}{1996}).

\bibitem[{\citenamefont{Sasaki and Tagoshi}(2003)}]{Sasaki:2003xr}
\bibinfo{author}{\bibfnamefont{M.}~\bibnamefont{Sasaki}} \bibnamefont{and}
  \bibinfo{author}{\bibfnamefont{H.}~\bibnamefont{Tagoshi}},
  \bibinfo{journal}{Living Rev. Rel.} \textbf{\bibinfo{volume}{6}},
  \bibinfo{pages}{6} (\bibinfo{year}{2003}), \eprint{gr-qc/0306120}.

\bibitem[{\citenamefont{Teukolsky}(1973)}]{Teukolsky:1973ha}
\bibinfo{author}{\bibfnamefont{S.~A.} \bibnamefont{Teukolsky}},
  \bibinfo{journal}{Astrophys. J.} \textbf{\bibinfo{volume}{185}},
  \bibinfo{pages}{635} (\bibinfo{year}{1973}).

\bibitem[{\citenamefont{Berti et~al.}(2006{\natexlab{a}})\citenamefont{Berti,
  Cardoso, and Casals}}]{Berti:2005gp}
\bibinfo{author}{\bibfnamefont{E.}~\bibnamefont{Berti}},
  \bibinfo{author}{\bibfnamefont{V.}~\bibnamefont{Cardoso}}, \bibnamefont{and}
  \bibinfo{author}{\bibfnamefont{M.}~\bibnamefont{Casals}},
  \bibinfo{journal}{Phys. Rev.} \textbf{\bibinfo{volume}{D73}},
  \bibinfo{pages}{024013} (\bibinfo{year}{2006}{\natexlab{a}}),
  \eprint{gr-qc/0511111}.

\bibitem[{\citenamefont{Berti et~al.}(2006{\natexlab{b}})\citenamefont{Berti,
  Cardoso, and Casals}}]{berti2006erratum}
\bibinfo{author}{\bibfnamefont{E.}~\bibnamefont{Berti}},
  \bibinfo{author}{\bibfnamefont{V.}~\bibnamefont{Cardoso}}, \bibnamefont{and}
  \bibinfo{author}{\bibfnamefont{M.}~\bibnamefont{Casals}},
  \bibinfo{journal}{Physical Review D} \textbf{\bibinfo{volume}{73}},
  \bibinfo{pages}{109902} (\bibinfo{year}{2006}{\natexlab{b}}).

\bibitem[{\citenamefont{Barrowes et~al.}(2004)\citenamefont{Barrowes, O'Neill,
  M., and Kong}}]{BONGK:2004}
\bibinfo{author}{\bibfnamefont{B.~E.} \bibnamefont{Barrowes}},
  \bibinfo{author}{\bibfnamefont{K.}~\bibnamefont{O'Neill}},
  \bibinfo{author}{\bibfnamefont{G.~T.} \bibnamefont{M.}}, \bibnamefont{and}
  \bibinfo{author}{\bibfnamefont{J.~A.} \bibnamefont{Kong}},
  \bibinfo{journal}{Studies in Applied Mathematics}
  \textbf{\bibinfo{volume}{113}}, \bibinfo{pages}{271} (\bibinfo{year}{2004}).

\bibitem[{\citenamefont{Doetsch}(1974)}]{Doetsch1974}
\bibinfo{author}{\bibfnamefont{G.}~\bibnamefont{Doetsch}},
  \emph{\bibinfo{title}{Introduction to the Theory and Application of the
  Laplace Transformation}} (\bibinfo{publisher}{Springer Berlin Heidelberg},
  \bibinfo{address}{Berlin, Heidelberg}, \bibinfo{year}{1974}), ISBN
  \bibinfo{isbn}{978-3-642-65690-3},
  \urlprefix\url{http://dx.doi.org/10.1007/978-3-642-65690-3_36}.

\bibitem[{\citenamefont{Ching et~al.}(1995)\citenamefont{Ching, Leung, Suen,
  and Young}}]{Ching:1995tj}
\bibinfo{author}{\bibfnamefont{E.~S.~C.} \bibnamefont{Ching}},
  \bibinfo{author}{\bibfnamefont{P.~T.} \bibnamefont{Leung}},
  \bibinfo{author}{\bibfnamefont{W.~M.} \bibnamefont{Suen}}, \bibnamefont{and}
  \bibinfo{author}{\bibfnamefont{K.}~\bibnamefont{Young}},
  \bibinfo{journal}{Phys. Rev.} \textbf{\bibinfo{volume}{D52}},
  \bibinfo{pages}{2118} (\bibinfo{year}{1995}), \eprint{gr-qc/9507035}.

\bibitem[{{\relax DLMF}()}]{NIST:DLMF}
{\relax DLMF}, \emph{\bibinfo{title}{{NIST Digital Library of Mathematical
  Functions}}}, \bibinfo{howpublished}{http://dlmf.nist.gov/, Release 1.0.5 of
  2012-10-01}, \bibinfo{note}{online companion to \cite{Olver:2010:NHMF}},
  \urlprefix\url{http://dlmf.nist.gov/}.

\bibitem[{\citenamefont{Leaver}(1986{\natexlab{b}})}]{Leaver:1986a}
\bibinfo{author}{\bibfnamefont{E.~W.} \bibnamefont{Leaver}},
  \bibinfo{journal}{J.\ Math.\ Phys.} \textbf{\bibinfo{volume}{27}},
  \bibinfo{pages}{1238} (\bibinfo{year}{1986}{\natexlab{b}}).

\bibitem[{\citenamefont{Abramowitz and Stegun}(1972)}]{bk:AS}
\bibinfo{author}{\bibfnamefont{M.}~\bibnamefont{Abramowitz}} \bibnamefont{and}
  \bibinfo{author}{\bibfnamefont{I.}~\bibnamefont{Stegun}},
  \emph{\bibinfo{title}{Handbook of Mathematical Functions}}
  (\bibinfo{publisher}{Dover Publications}, \bibinfo{year}{1972}).

\bibitem[{\citenamefont{Leung et~al.}(2003)\citenamefont{Leung, Maassen van~den
  Brink, Mak, and Young}}]{Leung:2003ix}
\bibinfo{author}{\bibfnamefont{P.~T.} \bibnamefont{Leung}},
  \bibinfo{author}{\bibfnamefont{A.}~\bibnamefont{Maassen van~den Brink}},
  \bibinfo{author}{\bibfnamefont{K.~W.} \bibnamefont{Mak}}, \bibnamefont{and}
  \bibinfo{author}{\bibfnamefont{K.}~\bibnamefont{Young}}
  (\bibinfo{year}{2003}), \eprint{gr-qc/0307024}.

\bibitem[{\citenamefont{Kavanagh et~al.}(2016)\citenamefont{Kavanagh, Ottewill,
  and Wardell}}]{Kavanagh:2016idg}
\bibinfo{author}{\bibfnamefont{C.}~\bibnamefont{Kavanagh}},
  \bibinfo{author}{\bibfnamefont{A.~C.} \bibnamefont{Ottewill}},
  \bibnamefont{and} \bibinfo{author}{\bibfnamefont{B.}~\bibnamefont{Wardell}},
  \bibinfo{journal}{Phys. Rev.} \textbf{\bibinfo{volume}{D93}},
  \bibinfo{pages}{124038} (\bibinfo{year}{2016}), \eprint{1601.03394}.

\bibitem[{\citenamefont{Castro et~al.}(2013)\citenamefont{Castro, Lapan,
  Maloney, and Rodriguez}}]{castro2013black}
\bibinfo{author}{\bibfnamefont{A.}~\bibnamefont{Castro}},
  \bibinfo{author}{\bibfnamefont{J.~M.} \bibnamefont{Lapan}},
  \bibinfo{author}{\bibfnamefont{A.}~\bibnamefont{Maloney}}, \bibnamefont{and}
  \bibinfo{author}{\bibfnamefont{M.~J.} \bibnamefont{Rodriguez}},
  \bibinfo{journal}{Classical and Quantum Gravity}
  \textbf{\bibinfo{volume}{30}}, \bibinfo{pages}{165005}
  (\bibinfo{year}{2013}).

\bibitem[{\citenamefont{Abbott et~al.}(2016{\natexlab{b}})}]{Abbott:2016nmj}
\bibinfo{author}{\bibfnamefont{B.~P.} \bibnamefont{Abbott}}
  \bibnamefont{et~al.} (\bibinfo{collaboration}{Virgo, LIGO Scientific}),
  \bibinfo{journal}{Phys. Rev. Lett.} \textbf{\bibinfo{volume}{116}},
  \bibinfo{pages}{241103} (\bibinfo{year}{2016}{\natexlab{b}}),
  \eprint{1606.04855}.

\bibitem[{\citenamefont{Olver et~al.}(2010)\citenamefont{Olver, Lozier,
  Boisvert, and Clark}}]{Olver:2010:NHMF}
\bibinfo{editor}{\bibfnamefont{F.~W.~J.} \bibnamefont{Olver}},
  \bibinfo{editor}{\bibfnamefont{D.~W.} \bibnamefont{Lozier}},
  \bibinfo{editor}{\bibfnamefont{R.~F.} \bibnamefont{Boisvert}},
  \bibnamefont{and} \bibinfo{editor}{\bibfnamefont{C.~W.} \bibnamefont{Clark}},
  eds., \emph{\bibinfo{title}{{NIST Handbook of Mathematical Functions}}}
  (\bibinfo{publisher}{Cambridge University Press}, \bibinfo{address}{New York,
  NY}, \bibinfo{year}{2010}), \bibinfo{note}{print companion to
  \cite{NIST:DLMF}}.

\end{thebibliography}

\bibliographystyle{apsrev}

\end{document}